%% file: main.tex
\documentclass[11pt]{article}

\usepackage{geometry}
\usepackage{fullpage}
\usepackage{graphicx}
\usepackage{pdflscape}
\usepackage{multirow}
\usepackage{ragged2e}

\usepackage{dcolumn}
\usepackage{caption}
\usepackage{booktabs,calc}
\usepackage{array}
\usepackage{tabularx}
\usepackage{paralist}
\usepackage{verbatim}
\usepackage{mathrsfs}
\usepackage{subfig}
\usepackage{setspace}
\usepackage{natbib}
\usepackage{amsmath}
\usepackage{amssymb}
\usepackage{amsthm}
\usepackage{enumerate}
\usepackage[linktocpage=true, colorlinks=true, linkcolor=blue, citecolor=black]{hyperref}
\usepackage{placeins}
\usepackage{authblk}
\usepackage{multicol}
\usepackage{array}
\usepackage[usenames, dvipsnames]{xcolor}

\newcolumntype{J}[1]{>{\justifying\arraybackslash}p{#1}}
\def\sym#1{\ifmmode^{#1}\else\(^{#1}\)\fi}
\newcolumntype{C}[1]{>{\centering\let\newline\\\arraybackslash\hspace{0pt}}m{#1}}

\usepackage[toc,page]{appendix} 

\begin{document}

	\title{Food Without Fire: Nutritional and Environmental Impacts from a Solar Stove Field Experiment\thanks{Corresponding author email: jdmichler@arizona.edu. We owe a particular debt of gratitude to Crosby Menzies and SunFire for their help in setting up the experiment and co-funding the cost of the solar stove training day. We wish to acknowledge the Zambian Ministries of Agriculture and Health along with the Barotse Royal Establishment for their support in conducting the RCT. Verena Nowak and Jessica Raneri provided helpful comments and suggestions in designing the experiment. Mulele Sibeso assited in translating the food diaries into English, which he provide particularly adept at, given the prevelance of local, non-standard names to food items. We appreciate the work of Agatha Changau, Bukwa Nyoni, Cynthiah Mpofu, Emil Kee-Tui, Grace  Kairezi, Hlozokuhle Mlilo, and Simelokuhle Moyo for transcribing the food diaries and of Nebai Hernandez for validating the transcribed data. Vanessa Ocampo provided assistance in preparing the data for analysis. Mary Arends-Kuenning, Anna Josephson, and participants at the 2021 AAEA \& WAEA Joint Annual Meeting in Austin provided helpful comments. We acknowledge financial support from the University of Illinois [Grant number: 2015-03766-03], the SPIA program ``Strengthening Impact Assessment in the CGIAR System (SIAC),'' and the CGIAR research programs on Aquatic Agricultural Systems and on Agriculture for Nutrition and Health. The experiment was registered at the AEA RCT registry, ID AEARCTR-0004054, and received IRB approval from the University of Illinois (\#16539).}}
	
	\author[a]{Laura E. McCann}
	\author[a]{Jeffrey D. Michler}
	\author[b]{Maybin Mwangala}
        \author[a]{Osaretin Olurotimi}
	\author[c]{Natalia Estrada Carmona}
	\affil[a]{\small \emph{University of Arizona, Tucson, USA}}
	\affil[c]{\small \emph{Ministry of Health of Zambia, Kaoma, Zambia}}
	\affil[c]{\small \emph{Bioversity International, Montpellier, France}}
	
	\date{}
	\maketitle
	
	\thispagestyle{empty}
	
\begin{center}\begin{abstract}
\noindent Population pressure is speeding the rate of deforestation in Sub-Saharan Africa, increasing the cost of biomass cooking fuel, which over 80\% of the population relies upon. Higher energy input costs for meal preparation command a larger portion of household spending which in turn induces families to focus their diet on quick cooking staples. We use a field experiment in Zambia to investigate the impact of solar cook stoves on meal preparation choices and expenditures on biomass fuel. Participants kept a detailed food diary recording every ingredient and fuel source used in preparing every dish at every meal for every day during the six weeks of the experiment. This produces a data set of 93,606 ingredients used in the preparation of 30,314 dishes. While treated households used the solar stoves to prepare around 40\% of their dishes, the solar stove treatment did not significantly increase measures of nutritional diversity nor did treated households increase the number of dishes per meal or reduce the number of meals they skipped. However, treated households significantly reduced the amount of time and money spent on obtaining fuel for cooking. These results suggest that solar stoves, while not changing a household’s dietary composition, does relax cooking fuel constraints, allowing households to prepare more meals by reducing the share of household expenditure that goes to meal preparation.
\end{abstract}\end{center}

{\small \noindent\emph{JEL Classification}: C93; D12; O13; Q10 
	}

 \setcounter{page}{0}
 \thispagestyle{empty}

\newpage
\onehalfspacing
\section{Introduction}

In rural Sub-Saharan Africa, over 80\% of households rely on solid fuels (e.g., wood, animal dung, agricultural waste, charcoal, and coal) as the primary source of energy for cooking their meals \citep{riva2017design}. Reliance on woodfuel and increasing population pressure in many parts of Continent has expedited deforestation and, by extension, negatively impacted human health. In the Barotse Floodplain of Western Zambia, the forested landscape is estimated to have diminished by 2\% annually from 2001-2014. As this rapid exploitation continues, the opportunity cost of collecting firewood or purchasing charcoal increases. Household members must venture farther from habited areas to find wood or pay higher prices to charcoal producers. With climbing prices of energy inputs for meal preparation, households are less able to meet their nutritional needs, as they choose to prepare a limited variety of quick-cooking foods, such as corn meal and leafy greens \citep{barbieri2017cooking}. Evidence suggests that less diverse diets negatively impact several broader health outcomes, including birthweight and blood pressure \citep{ruel2003operationalizing, pangaribowo2013food}.

In this paper, we investigate the link between rising fuel costs for cooking and nutritional and environmental outcomes. We conduct a field experiment in the Barotse region of Zambia and examine the impact of provisioning a household with a solar cook stove on household meal preparation and expenditure on fuel for cooking. Over the length of the experiment, treatment and control households recorded the ingredients used in each dish cooked for each meal every day. This provides us with a database of 93,606 ingredients used to prepare 30,314 dishes, an average of 195 dishes per household. We also have a weekly record of the time and money households spent procuring fuel for cooking.

We find that households assigned solar stoves used them to prepare 40\% of the dishes consumed throughout the experimental period. We find no significant effect of solar stove assignment on measures of household dietary diversity or the frequency of cooking. Regarding fuel use, we find significant and economically meaningful effects. On average, households spent 45 fewer minutes a week collecting firewood and saved, on average, 22 USD in fuel costs. The savings in fuel costs come from a reduction in the use of biomass fuels like firewood and charcoal. This cost savings translates to potential annual savings of 182 USD. Given that the stoves cost 90 USD and have a life expectancy of $10$ years, this makes them highly cost-effective.

This paper contributes to three streams in the literature. First, we contribute to the economic health and nutrition literature in terms of our approach to data collection. Food diaries have long been used in the nutritional science literature as the most accurate method for capturing food intake \citep{WisemanEtAl05}. However, in health and development economics, the most common way to collect food consumption data has for many years been to ask respondents to recall ingredients that went into meals consumed over the last 24 hours \citep{StraussThomas98}. While researchers recognize that a single 24-hour recall is likely to yield noisy estimates of average food consumption, since there is substantial variation in day-to-day eating habits, few studies collect multiple recalls over time or implement diaries, given the added costs of data collection \citep{FiedlerEtAl12}. There have been several studies that examine the trade-offs between 24-hour recall and food diaries \citep{Gibson02, SmithEtAl06, ZEZZA20171}. These studies tend to find that over a short period (three to ten days), diaries appear more accurate than recall \citep{Gibson02, SmithEtAl06, BeegleEtAl12, BACKINYYETNA20177, BRZOZOWSKI201753, CONFORTI201743}, though fatigue in completing diaries makes them less reliable when diary length extends past ten days \citep{TROUBAT2017132}. Our diaries last six weeks (42 days), which, to our knowledge, is the longest food diary study in economics. We combat fatigue by incentivizing participants with an in-kind reward upon diary completion.\footnote{We also have participants explicitly mark when entries were completed at the time of cooking or if an entry was completed after the fact (recall).} We find no evidence of fatigue in our data.

Second, we contribute to the literature that quantifies the impact of adopting relatively green energy technology on household-induced deforestation and environmental degradation, typically measured by reductions in biomass use. Although studies have documented that improved cookstoves result in firewood reduction \citep{frapolli,BENSCH2015187, JEULAND2020149,  MEKONNEN2022107467, wentzel2007} and time savings \citep{afridi}, most of these studies have come from improved cook stoves in general, which may be biomass burning, and not solar stoves in particular. Studies have also suggested that there might be limited firewood savings from adopting solar stoves because households may be participating in energy stacking, i.e., using different cooking sources as complements rather than substituting cleaner technology for biomass fuels \citep{BAUER2016250}. Another reason for why some studies find no impact of fuelwood use is because of inadequate solar stove designs for the local context \citep{beltramo}. In this paper, we find no evidence of stacking but rather substantial savings in firewood and biomass fuel in both time saved and quantities used.

Finally, we contribute to the literature that studies the health and nutritional impact of solar and other improved or clean cook stoves. While solar stove technology has existed for centuries, improved solar stoves entered the development discourse in the latter half of the twentieth century \citep{biermann1999solar, wentzel2007}. Experiments for evaluating the impacts of solar stoves, as well as other improved cooking stoves, became more popular following the endorsement by the United Nations through their Sustainability for All plan aimed towards increasing the adoption of improved cook stoves \citep{bensch2015intensive, iessa2017s}. Solar stoves have been presented as a solution to numerous development problems, such as deforestation, respiratory and nutritional health, and limited financial resources. However, many studies find low adoption or low willingness-to-pay for solar and improved cook stoves \citep{biermann1999solar, ruiz2011adoption, mobarak2012low, hanna2016up, iessa2017s, bensch2020}. While the majority of the literature finds low rates of adoption, there are instances of success. \cite{bensch2015intensive} found high adoption rates, substantial improvements to respiratory health, and substantial savings in firewood consumption. Most of the impact evaluation literature is limited to stove impact on respiratory conditions \citep{hanna2016up, smith2004reducing, smith2009, mobarak2012low, iessa2017s}. We expand the scope of this literature by investigating dietary and nutritional impacts in addition to impacts on biomass fuel consumption.

\section{Study Context and Design}

\subsection{The Barotse Floodplain System}
Spanning roughly 230 km, the Barotse Floodplain System (BFS) includes four districts in Zambia's Western Province and is home to an estimated 225,000 residents \citep{emerton2003barotse,zimba2018assessment}. The BFS is inundated annually from December to March from the flooding of the Upper Zambezi River. This yearly event revitalizes the soil and facilitates the agricultural livelihoods of 90\% of the individuals in the region \citep{pasqualino2016food, baars2001grazing, turpie1999economic, mirriam2019climate, flint2008socio}.

The Lozi people are the largest ethnic group in the region and maintain oversight of the area’s agricultural operations through the Barotse Royal Establishment governing body \citep{pasqualino2016food}. The migratory patterns of the Lozi are adapted to the natural flow of the Zambezi River and move twice annually: to the wooded upland during the flooding season and to the grassy lowland when the water recedes \citep{baars2001grazing, pasqualino2016food, cai2017living, joffre2017increasing, rickertlearning}. The Lozi are primarily subsistence households earning their livelihoods by crop production, fishing, and cattle grazing \citep{pasqualino2016food}.

The BFS experiences severe poverty and is highly vulnerable to internal and external shocks \citep{flint2008socio, rajaratnam2015social}. Historically, floodplain environments have served an integral role in sustaining life by replenishing nutrients vital to aquatic and terrestrial systems. In recent decades, this seasonal dependence has evolved into what \cite{cole2018postharvest} characterize as a social-ecological trap, where rapid population growth and overexploitation of natural resources have created a cycle of deforestation, food insecurity, and poverty. The region experiences a period of four to five months with limited access to food \citep{castine2013increasing, baidu2014assessment, rajaratnam2015social}. These aforementioned ecological barriers are compounded by limited access to agricultural inputs, equipment, and household knowledge about improved management techniques \citep{baidu2014assessment}. 

From 2011 to 2015, World Fish and Bioversity International oversaw two development programs in the BFS: Aquatic and Agricultural Systems (AAS) and Agriculture for Nutrition and Health (ANH). Rresearchers and practitioners in these programs adopted and implemented systems research, participatory action research, and a gender transformative approach to identify locally relevant opportunities to overcome poverty and increase nutrition and food security while maintaining and improving natural resources and the ecosystem services these provide \citep{madzuzo2013, pasqualino2015market, pasqualino2015seasonal, pasqualino2016food}. In practice, these programs supported multiple and complementing activities that aimed to increase the knowledge, value, consumption, and production of diverse crops, notably traditional ones \citep{ObornEtAl17, DelRioEtAl18, EstradaCarmonaEtAl20}. The findings from these studies guide our experimental design.

For example, as part of the AAS program, households in the local communities self-selected into activities such as nutrition cooking clubs, diversified farmer learning plots, or both. The nutritional cooking clubs aimed to enhance nutritional knowledge and improve dietary habits and overall health while fostering creativity and a love for cooking. Additionally, these clubs sought to revive nutritious recipes and foods lost in contemporary diets. The goal of the diversified learning plots was to learn by doing and from peers regarding diversification options that can ensure food production all year round. Farmers experimented with different crops and different sustainable practices such as rotation, cover crops or heavy mulching, intercropping, no burning, agroforestry, and no-till. Households could participate in one or both of the groups or refrain from participating in any of the programs.

\subsection{Sample Selection and Intervention}

Ten villages across the BFS participated in the AAS program as AAS Communities \citep{pasqualino2016food}. Seasonal crop production and accessibility to markets, education, and health care vary across the villages and so the program activities vary by community \citep{pasqualino2015seasonal}. From the ten AAS communities, we selected three communities that had the presence of both nutrional cooking clubs and diversified learning plots. We stratified our random sample by village because of demographic and population differences between villages.

To account for participants' previous exposure to the AAS development programs, we further stratified our sample by household involvement in these activities. This strategy resulted in four sub-samples, including i) households that participated in learning plots; ii) households that participated in cooking clubs; iii) households that participated in both learning plots and cooking clubs; and iv) households that participated in neither. In our analysis we include strata fixed effects for both village and AAS group. 

The intervention took place in March and April 2016. In each community, we facilitated an introductory day-long event. During the morning, we had an open discussion with participants about the objectives of the study, the length of the study, the time commitment, and the expected results from the project. Next, we conducted a demonstration of the solar stoves to make a communal meal, highlighting safety management and precautionary measures with hands-on experience. During the afternoon session, we collected the names of attendees interested in participating in the study. From the subset of interested, consenting participants, we randomly drew names by strata, without replacement, for the assignment of the solar stoves. Across the three villages, 61 households were assigned to treatment (received a solar stove), and 96 were assigned to control for a total sample size of 157 households.

\begin{figure}[!htbp]
	\begin{minipage}{\linewidth}		
		\caption{Solar Stove Demonstration}
		\label{fig:solar_stove}
		\begin{center}
			\includegraphics[width=.75\linewidth,keepaspectratio]{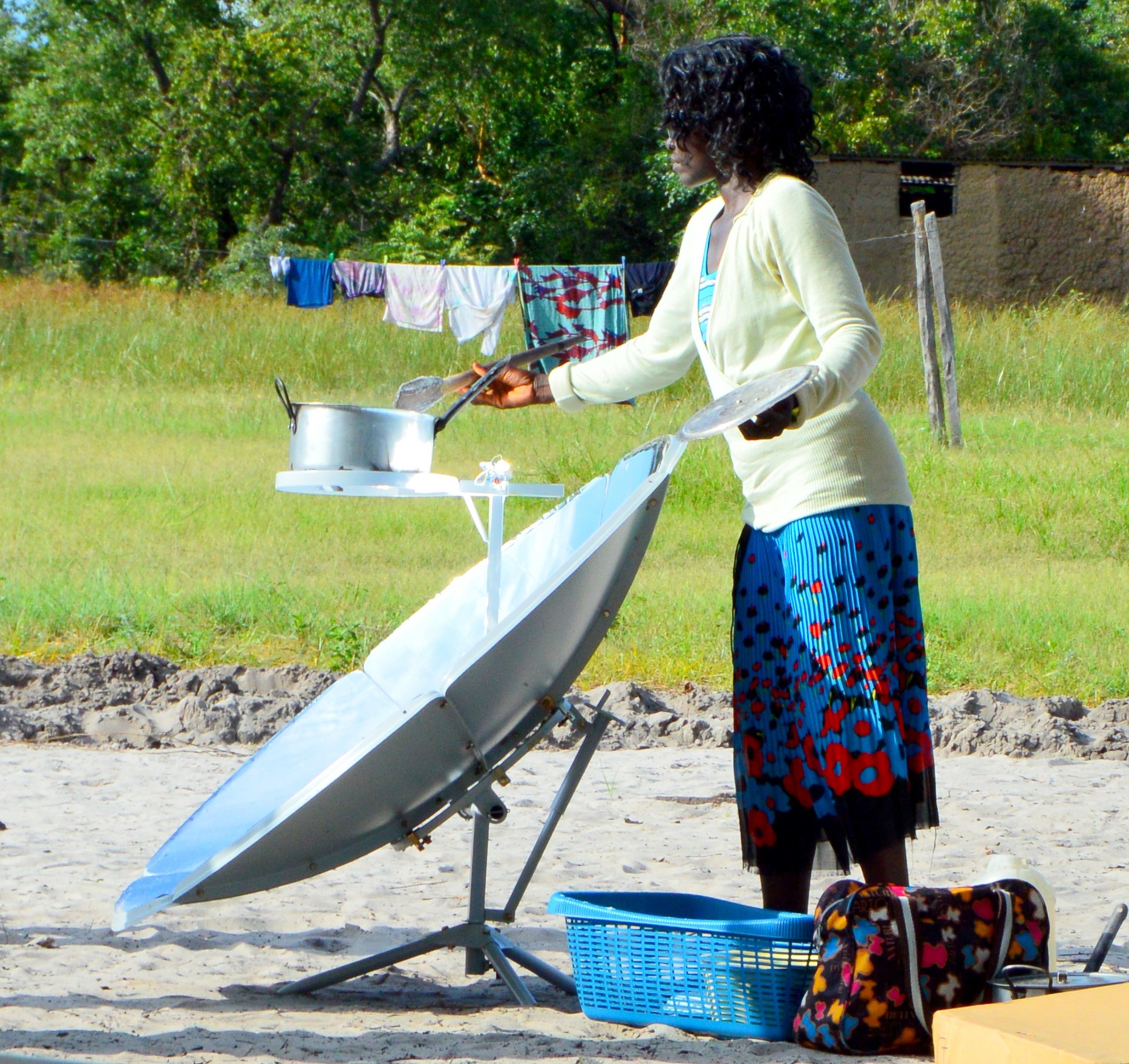}
		\end{center}
		\footnotesize  \textit{Note}: Image is of one of the SunFire solar stoves used in the field experiment. The photo comes from one of the introductory day-long events the team held to familarize potential participants prior to random assignment and the start of the experiment.
	\end{minipage}	
\end{figure}

The solar stoves themselves are Sunfire Solution's SunFire12 Parabolic Solar Cookers. The stoves are 1.2 meters in diameter and focus solar energy on a single heat area at the bottom of a pot. Figure~\ref{fig:solar_stove} provides an example of the solar stove. The reflective surface is created by metallicized tape that can be replaced as it becomes worn or damaged. This extends the life of the stove to between 10 and 12 years. The stoves retail for about 85 USD. While this is a significant up-front cost for households in the BFS, the annualized cost is around eight USD.

A unique element of our study is our approach to data collection. In the morning of our day-long event, we explained to potential participants that they would be recording/collecting the data themselves by maintaining a detailed food diary.\footnote{See Online Appendix \ref{sec:app_b} for a reproduction of the food diary tool.} The diary asked households to record every ingredient used in each dish for every meal eaten every day over the six weeks of the experiment. We also asked them to record the source of heat for preparing each dish (solar stove, firewood, charcoal, cow dung), along with how many times they boiled water or milk during the day. Finally, we asked participants to, at the end of each week, to record the amount of time or money spent in obtaining firewood, charcoal, or cow dung. To incentivize compliance among control households, every participating household was entered into a raffle of the solar stoves at the end of the six-week experiment, conditional on the satisfactory completion of the food diary. Those who did not win the raffle received a 500 ml bottle of oil and 250 g of local seeds as a contribution to their time and effort in collecting the data. 

\section{Data}

\subsection{Recording and Transcribing the Diaries}

Our approach to data collection, which produces rich and extensive data on food consumption for our study, also creates unique challenges, or at least unique given that most fieldwork now uses computer-assisted personnal interview software. After the experiment, we collected the paper food diaries maintained by all 157 households. These were scanned as images and then individually transcribed into a spreadsheet. This required deciphering individual handwriting in the local Lozi language. Data transcription was then validated by an individual not involved in the original transcription process, who also checked for obvious misspellings, case sensitivity, and errant characters from data entry. We then created a Lozi-English food dictionary, which was not always straightforward, given the prevalence of local, non-standard names for plants and animal protein used in cooking. Finally, we determined the scientific species name from which each ingredient came.

The resulting food consumption data set contains 93,606 ingredients used to prepare 30,314 dishes eaten as part of 15,899 meals by the 157 families over the six weeks. We identify 111 unique ingredients from 63 unique species. There are more ingredients than species since things like maize flour and fresh maize or pumpkin flesh and pumpkin leaves are unique ingredients but not unique species. We further categorize ingredients into 12 food groups following The Food and Agriculture Organization of the United Nations (FAO) guidelines \citep{faoguide}.\footnote{See Online Appendix \ref{sec:app_b} for tables summarizing the frequency and types of ingredients along with the frequency of species and food groups.}

An obvious concern with the data is that the length of time households were asked to maintain the food diaries will result in fatigue and a loss of accuracy in record keeping. Previous research has shown that data quality in food diaries declines if diary length runs more than ten days \citep{TROUBAT2017132}. Our diaries cover 42 days. To minimize fatigue, we provided incentives to households, as described above. We also verify that changes in the data, such as number of ingredients recorded or number of meals skipped, are not correlated with time or treatment status.\footnote{See Online Appendix~\ref{sec:app_b} for discussion of these results.} We find no evidence that data quality deteriorated over time in our study.

\subsection{Dietary Composition}

We are interested in whether the provisioning of solar stoves, by lowering the cost of cooking, changes the dietary composition of the household. We measure dietary composition in three different ways.\footnote{See Online Appendix~\ref{app:summary} for summary statistics of each dietary composition measure.} First, we use the FAO's Household Dietary Diversity Score (HDDS), which categorizes ingredients into one of 12 FAO-designated food group categories. The HDDS is simply the count of the unique food group categories present in a given dish, meal, day, etc., where one indicates low dietary diversity, and 12 indicates high dietary diversity.

Second, following \cite{lachat2018dietary}, we include dietary species richness (DSR) as a measure of dietary diversity. The DSR is calculated by matching ingredients to species and tabulating the number of unique species consumed during a given period. \cite{lachat2018dietary} highlight that dietary species richness serves as an important validation tool and enriches the understanding of nutrient availability and biodiversity for a given population. While HDDS varies between one and 12, DSR has no upper bound since it counts the number of unique species present in a given dish, meal, day, etc. In our data, 63 unique species were consumed over the length of the study.

Third, we include a measure of legume consumption by the household. One component of the AAS program was to encourage households to grow and consume more legumes. Legumes are viewed as an important component of sustainable or climate-smart agriculture. They fix nitrogen in the soil, reducing the need to add nitrogen to the soil through inorganic fertilizer. Legumes are also protein-rich, so adding them to one's diet can offset reliance on animal-based protein. One hurdle to adopting legumes is their long cooking time, which can make preparing them prohibitively expensive if one must purchase fuelwood for cooking. Our measure of legume consumption is simply counting the number of times legumes were present in a given dish, meal, day, etc.

\subsection{Cooking Frequency}

In addition to changes to the make-up of a household's diet, we are also interested in the frequency with which a household prepares dishes or meals. Because food habits are cultural and historical, they may not change in a discernable way over the length of our experiment. But, by lowering the cost of meal preparation, household may prepare more dishes in a meal or eat more meals (skip fewer meals). We measure cooking frequency in two ways. First, we count of the number of dishes prepared in a given day, week, or over the entire length of the study. The second measure of cooking frequency is counting up the number of meals a household skipped in a day, week, or during the total study period.

\subsection{Fuel Collection and Purchases}

In addition to changing household eating habits, we expect the provisioning of solar stoves to have environmental impacts through a reduction in the collection of firewood and/or the purchase of charcoal. The food diaries asked households to record the amount of time individuals spent each week collecting firewood and/or cow dung for use as cooking fuel. The diaries also asked households to record the number of times individuals purchased firewood or charcoal for cooking. In general, nearly all data on collection was for firewood (not cow dung) and all data for purchases was for charcoal (not firewood). Households were also asked the quantity collected/pruchased and the price of purchases. We measure environmental impacts in four ways: the amount of time spent collecting firewood, the quantity of firewood collected or purchased, the quantity of charcoal purchased, and the total value in USD for fuel used by the household. We value collected fuel at market prices.

\subsection{Controls}

We did not collect baseline data and, therefore, did not conduct balance tests across the control and treatment groups. Rather, we collected household characteristics that are likely to be time-invariant while the experiment was ongoing. During the experiment, we conducted periodic check-ins to ensure compliance with treatment assignment and maintenance of the diaries. During these visits, a team member collected gender, age, and education data at the individual level. They also captured household-level data on the number of household members and information on asset ownership. On average, participants are in their late 40s with low education and asset ownership levels. Our sample includes more females than males, and households have seven household members on average.

Since solar stoves are most efficient when it is sunny, we also include a measure of cloud cover to control for weather events that might limit the ability to use a solar stove. We accessed Landsat 8 data to calculate the percentage of cloud cover during the study period.\footnote{See Online Appendix \ref{app:summary} for details about resolution and coverage in Landsat 8 along with summary statistics of cloud cover and of the other control variables.} It is important to note that these data are incomplete, as the satellite was only above the region five times during the experiment. Thus, we use the single value obtained from Landsat 8 for each day within the week. When no data were available for a particular week, we assigned it the averaged values from the week before and the week following.

\section{Empirical Framework}

\subsection{Causal Pathways and Outcomes}

Before outlining our empirical strategy, it is useful to explicitly state the assumptions underlying the causal pathway from the provision of solar stoves to measured outcomes. First, we assume that the behavior of all households in the study can be approximated using the agricultural household model \citep{deJanvryEtAl91}. Essentially, households are maximizing their utility of consumption while also optimizing production and facing binding budget constraints that forces them to make trade-offs. Second, we assume that solar cook stoves are effective at providing a way to prepare meals at zero cost in terms of fuel to heat the food (there is still labor cost in cooking). Third, we assume that households provided with a solar stove use the solar stove. Finally, we assume that relaxing the budget constraint by changing the cost of meal preparation results in both income and substitution effects such that households would change their behavior along three fronts: (1) the ingredient make-up of dishes; (2) the frequency with which they prepare dishes; and (3) the effort or cost spent on obtaining fuel for cooking. Based on these assumptions, the causal pathway is straightforward. Solar stoves will lower the cost of meal preparation in such a way that households using the stoves will change their behavior around cooking practices. This change will manifest itself in a different ingredient make-up in the dishes, a change in the frequency of dish preparation, and a change in time and monetary expenditure on fuel.

The above assumptions are sufficient to outline a causal pathway for reallocation of effort among households (change the price, change the consumption bundle). If we add one additional assumption, we can sign the direction of these changes. A fifth assumption that we make is that in optimizing under constraints, households make sub-optimal dietary and nutrition choices relative to what they would make if they had more money. Essentially, we are assuming that households will not just reallocate behavior around cooking but reallocate it in such a way as to improve their diet (both on the intensive and extensive margin). If that assumption is true, then we expect to see (1) an increase in the nutritional content of the dishes they prepare, (2) an increase in the frequency of food preparation, and (3) a decrease in the expenditure of cooking fuel.\footnote{It is possible that a household would not change fuel expenditures but rather hold them constant and use the solar stove to further increase the frequency of meal preparation.}

Given the causal pathway, our primary outcomes of interest are the make-up of dishes households prepare and the frequency with which households prepare dishes. Additionally, we report on the frequency of solar stove use (an intermediate outcome) and changes in fuel expenditure (a secondary outcome).

\subsection{Treatment Effect Estimation}

We start by examining the intermediate outcome of solar stove use. We estimate the average treatment effect (ATE) of being given a solar stove on the use of the solar stove as follows: 

    \begin{equation}
            S_{ht} = \alpha + \beta T_{h} + X_{h}' \gamma + \mu_{v} + \phi_{g} + \epsilon_{iht}. \label{eq:ate}
    \end{equation}

\noindent Here, our outcome variable $S_{ht}$ represents the frequency of solar stove use in preparing a dish for each household \textit{h} during time \textit{t}. \textit{T} is an indicator variable denoting assignment to solar stove use; \textit{X} is a matrix of covariates indexed by household; $\mu$ is a village fixed effect indexed by village \textit{v}; $\phi$ is a strata fixed effect based on enrollment in AAS activities; and $\epsilon$ is the stochastic error term for each household which, due to the randomization, is orthoganol to the treatment. 

For all other outcomes (primary nutritional outcomes; secondary fuel outcomes), we estimate the intent-to-treat (ITT) effects with the following model:

\begin{equation}
        Y_{ht} = \alpha + \beta T_{h} + X_{h}' \gamma +\mu_{v} + \phi_{g} + \epsilon_{ht}. \label{eq:itt}
    \end{equation}

\noindent Here $Y_{ht}$ represents the outcome for household \textit{h} during time \textit{t}. All other variables are as defined in Equation~\eqref{eq:ate}.

Each of the models above is estimated at a variety of levels of aggregation. The primary unit of observation in the analysis is a dish, which is composed of a variety of ingredients and is cooked using a single heat source: solar stove or a type of biomass fuel.\footnote{See the food diary in Online Appendix \ref{sec:app_b} for more information on the structure of the data.} We can then aggregate dishes into meals (breakfast, lunch, or dinner), and meals into days, days into weeks, and weeks into a single aggregate value for the household for the entire six weeks. We report results both with and without controls. These controls include the age, gender, and educational level of the head of the household, the household size, an index of household assets, an index of livestock assets, and cloud cover for the week.

When possible, we use Liang-Zeger cluster-robust standard errors at the unit of randomization: the household. This corrects for heteroskedasticity and serial correlation when we have multiple observations for a household. However, when the outcome is aggregated to a single value for the entire six week period, we only have one observation per household. In these cases, we implement Eicker-Huber-White robust standard errors to account for heteroskedasticity, since serial correlation, by construction, is no longer present.

\section{Results}

We present results on the intermediate outcome of stove adoption, followed by results on our main outcomes of interest: dietary composition, cooking frequency, and fuel use.

Before considering specific outcomes of interest, it is important to understand how the treated respond to their treatment status. We do this by estimating the ATE of being given a solar stove on actually using a solar stove to preapre a dish. While being assigned a stove is random, the choice to use the stove to prepare a given dish is not.\footnote{As a robustness check, we use the difference between assignment and use to estimate the local average treatment effect (LATE) in Online Appendix~\ref{app:robust}.} The results in Table~\ref{tab:ss_use} suggest that participants assigned a solar stove use the stove to prepare approximately 40\% of dishes at the meal, day, week, and six-week level. This amounts to households using the stove about twice a day on average.

\subsection{Dietary Composition}

Next, we examine the effect of being assigned a solar stove on changes in a household's dietary composition. We measure dietary composition in three ways: HDDS, DSR, and the frequency of cooking legumes. For each of our measures, we present ITT estimates with and without controls by dish, meal, day, week, and entire six week period. The results, as seen in Table~\ref{tab:diverse_out}, tell us that households assigned a solar stove did not significantly change the number of food groups, the number of species, or the number of legumes consumed. In fact, for HDDS and DSR, the majority of coefficients are negative, suggesting that being assigned a solar stove contributed to a small reduction in the diversity of dietary composition. However, the negative coefficients are neither statistically or economically significant. The mean HDDS for the control group in a day (HDDS typically uses a 24-hour recall) is 5.7, meaning the household consumes food from just under six food groups in a day. By contrast, treated households' daily HDDS is 5.5, a difference of less than two tenths of a food group. Similarly, small reductions are present for DSR, while similarly small increases are seen for legumes. Households in the control consume legumes about four times a week. Treatment households consume legumes 0.28 more times a week.

We interpret our results as true nulls given the large number of observations we have at the dish, meal, and day level. One potentail explanation for the null findings is that, given our experiment was conducted during the harvest season, households may have already made and acted upon their agricultural production decisions before participating in the RCT. Such decisions directly impact food availability for rural subsistence households and in local markets. Another potential explanation is that, given food choices are functions of culture and history, a six week window of study is insufficient to produce measureable change in dietary habits. Whatever the explanation, these results imply that solar stove's effects may not be visible in the short term or that lifting cooking fuel constraints may not be enough to change dietary patterns, especially if the intervention is not synchronous with the agricultural planting season.

\subsection{Cooking Frequency}

While the provisioning of solar stoves did not change dietary composition, it may have an effect on the extensive margin by increasing cooking frequency. We measure cooking frequency as either the number of dishes prepared in a meal or the number of meals skipped by a household. For each of our measures, we present estimates with and without controls by day, week, and overall. The results in Table~\ref{tab:freq_out} show that households assigned a solar stove did not significantly change their cooking frequency. In general, the coefficients are in the direction we would expect; coefficients on the number of dishes prepared are positive and coefficients on the number of meals skipped are negative. This suggests that households are preparing more dishes at each meal and skipping fewer meals. On average, households in the control prepare about five dishes a day. Households with solar stoves prepare 0.1 more dishes or about two percent more dishes a day. Over the length of the experiment treatment households only prepare three more dishes than control households. More economically and nutritionally meaningful effects are present for meals skipped. Treatment households skip about one less meal than control households, who skip on average 26 meals during the 42 days of the study.

It is more difficult to interpret the null findings for cooking frequency than the null results for dietary composition. Given that solar stoves provide another method of cooking that can augment cooking with fuel, and that using the stove is free of monetary costs, one would expect an increase in cooking frequency, particularly is households are energy stacking. Additionally, frequency of cooking is not tied to cultural or historical food preferences nor to the types of food available in the market. One potential explanation is that households remain budget constrained and so are unable to procure additional food to prepare with the solar stove.

\subsection{Fuel Collection and Purchases}

Finally, we examine the environmental impact of solar stove assignment in terms of biomass use averted. We measure fuel costs in four ways: time spent gathering firewood, the quantity of firewood and charcoal collected or purchased, and the total value of fuel used by the hosuehold. For each of our measures, we present ITT estimates with and without controls by week and entire six week period. As seen in Table~\ref{tab:fuel_out}, treated households significantly reduce their firewood use and their fuel expense. There is also a reduction in charcoal purchases, though this is not statistically significant. Compared to control households, treated households use about 29kg less of firewood per week (153 kg over the six weeks). This result is also economoically meaningful in that it is nearly a 40\% reduction (32\% over six weeks) in the quantity of firewood used relative to control households. In addition, households spend about 45 minutes less per week gathering firewood. The time use is especially important as it suggests relaxing household production time constraints that could be used for other activities, e.g., childrearing, farming, or leisure. 
 
While the dietary and cooking impacts of solar stoves are small to non-existent, the effects on fuel are nearly all statistically significant and economically meaningful in terms of environmental impact. Not only are households collecting or purchasing less firewood, but this reduction has a direct economic impact. Treated households spend about four USD less on fuel per week or about 21 USD less on cooking fuel for the study period. Taken together, these results show that solar stoves have potential as a tool for simultaneously reversing environmental degradation from biomass use while relaxing household time and budget constraints.

\section{Conclusion}

 This paper examines nutrition and environmental outcomes associated with solar stove use in a randomized controlled trial. We hypothesize that when the time and cost burdens of procuring cooking fuel are reduced by access to solar stoves, households will be better equipped to allocate time and money to source more diverse ingredients for meal preparation,. Based on this hypothesis, we randomly assign solar stoves to households who keep detailed food diaries for six weeks. Our analysis shows that households assigned solar stoves used them to cook roughly 40\% of their meals. However, we find no significant effect of solar stove assignment on measures of household dietary diversity. We also do not see a significant effect of the provisioning of solar stoves on changes in frequency of cooking. We do find that stoves have a stastically significant and economically meaningful impact on the quantity of firewood used, time spent in gathering firewood, and overall fuel costs.

 Our results show that lifting constraints to cooking fuel costs in the short term may not change households' dietary patterns and outcomes, as households likely face other binding resource constraints that affect dietary composition. Additionally, the results imply that food choices are sticky and households may not deviate from their dietary preferences. These results pose further questions about the tenor, timing, and size of dietary constraints, the removal of which would translate to improvements in household nutritional composition and food security. Beyond these nutritional concerns, our results demonstrate that solar stoves can reduce the need for biomass fuel and relax household budget constraints related to food preparation costs. This suggests that solar stoves offer a cost effective way to address deforestation and increasing fuel costs, at least at a micro-level.


\newpage
\singlespacing
\bibliographystyle{chicago}
\bibliography{main}


\newpage 
\FloatBarrier

\begin{landscape}
\begin{table}[!htbp]	\centering
    \caption{Share of Dishes Prepared Using Solar Stoves} \label{tab:ss_use}
	\scalebox{1}
	{ \setlength{\linewidth}{.2cm}\newcommand{\contents}
		{\input{tables/ss_use}}
	\setbox0=\hbox{\input{tables/fuel_late}}
    \setlength{\linewidth}{\wd0-2\tabcolsep-.25em}
    \input{tables/fuel_late}}
\end{table}
\end{landscape}

\begin{landscape}
\begin{table}[!htbp]	\centering
    \caption{ITT Estimates of Solar Stove Assignment on Dietary Composition} \label{tab:diverse_out}
	\scalebox{1}
	{ \setlength{\linewidth}{.2cm}\newcommand{\contents}
		{\input{tables/diverse_out.tex}}
	\setbox0=\hbox{\input{tables/fuel_late}}
    \setlength{\linewidth}{\wd0-2\tabcolsep-.25em}
    \input{tables/fuel_late}}
\end{table}
\end{landscape}

\begin{table}[!htbp]	\centering
    \caption{ITT Estimates of Solar Stove Assignment on Frequency of Cooking}  \label{tab:freq_out}
	\scalebox{.95}
	{ \setlength{\linewidth}{.2cm}\newcommand{\contents}
		{\input{tables/freq_out.tex}}
	\setbox0=\hbox{\input{tables/fuel_late}}
    \setlength{\linewidth}{\wd0-2\tabcolsep-.25em}
    \input{tables/fuel_late}}
\end{table}

\begin{table}[!htbp]	\centering
    \caption{ITT Estimates of Solar Stove Assignment on Fuel Collection}  \label{tab:fuel_out}
	\scalebox{.95}
	{ \setlength{\linewidth}{.2cm}\newcommand{\contents}
		{\input{tables/fuel_out.tex}}
	\setbox0=\hbox{\input{tables/fuel_late}}
    \setlength{\linewidth}{\wd0-2\tabcolsep-.25em}
    \input{tables/fuel_late}}
\end{table}


\clearpage
\newpage
\appendix
\onehalfspacing

\begin{center}
	\section*{Online-Only Appendix to ``Food Without Fire:  Nutritional Impacts from a Solar Stove Field Experiment''} \label{sec:app}
\end{center}


\section{Daily Food Diaries and Ingredient Lists\label{sec:app_b}}

In this appendix, we present a copy of the food diary as well as summary statistics regarding the overall characteristics of the ingredients recorded by participants. We also provide evidence of a lack of fatigue or attrition in diary keeping by treatment status.

Figures~\ref{fig:diary1} and~\ref{fig:diary3} present the food diary in the Lozi language. In the top left of every page is a place for the household to record the dates that the diary covers. Each day's diary has the same structure. For each meal (breakfast, lunch, dinner) households have space to record up to four dishes and include up to six ingredients per dish. Most households cooked only one dish per meal (52\%), though some meals included five dishes, with households recording the extra dishes in the margins. The modal number of ingredients in a dish was three (35\%), but some dishes included up to seven, with households recording extra ingredients in the margins. After the list of ingredients in a dish, there is a space for households to record the cooking method (solar stove, firewood, charcoal, or cow dung), and we include reference images at the top of each page. Finally, at the bottom of each dish is space to record the volume of legumes cooked and to note if the ingredients in the dish were based on recall memory instead of recorded at the moment.

To the right of the daily meal entry is a space to record the number of times the household boiled liquids (up to nine times a day). Households were asked to record if they boiled water or milk, the fuel source, and the liquid volume. We do not report on liquids in this paper, though we did pre-specify it in the analysis.

The first two pages of the diary are identical and cover days one to six in a week. The final page for the week included space to record food preparation data for day seven plus space to record fuel collection and purchase information for the week. The diary provides space to record the family member who collected the fuel, whether it was firewood or cow dung, how many times that family member went collecting, how much time was spent, and an estimate of the weight and the value. Space is also provided to record similar information about purchasing firewood or charcoal.

Next, we present data on the existence of fatigue or attrition when maintaining the diaries. The left hadn side panel of Figure~\ref{fig:response} presents a histogram of the number of households that recorded information in the diary on each day of the study. The study lasted 42 days, though the start date varied by village. The histogram breaks down the daily response rate by treatment and control. If every household recorded information in the diary every day of the study, then the distribution of data in Figure~\ref{fig:response} would be perfectly uniform. While the data is not perfectly uniform, it clearly approximates a uniform distribution very closely. While a greater number of households were likely to record data in the diary on the first day than on any other day, by the second day, the frequency of recording was fairly constant over time. More importantly for our analysis, we see no meaningful difference in the frequency of recording by treatment group. We conduct an event study style analysis to further demonstrate a lack of difference in diary keeping by treatment (see right-hand panel). For treatment and control, we regress the number of ingredients in a dish on an indicator for each day in the study, with the last day as the omitted category. There is a decline in the number of ingredients that households recorded in each dish during the third week. However, confidence intervals for treatment and control overlap for the entire length of the study. This suggests that while there was some fatigue during the third week, even in that week, as well as all other weeks, fatigue did not differ by treatment status. Unlike other studies that rely on food diaries, we see no consistent or sustained decline in recording keeping across time.

Tables~\ref{tab:ingredients},~\ref{tab:food_groups}, and~\ref{tab:species} present summary statistics regarding the raw ingredient data from the food diaries. Tables~\ref{tab:ingredients} reports on the 25 most frequently used ingredients in cooking. The food diaries record the use of 93,606 ingredients, 111 of which are unique. Salt is the most common ingredient, excluding water, making up 15\%  of the records. Pre-processed maize porridge is the next most common (11\%) followed by cooking oil (11\%), maize and tomato (9\%), and maize flour (7\%). Cooked maize, in some form, is the staple in Zambia, eaten at nearly every meal. Beyond maize porrdige, the following most frequent ingredients are vegetables. Meat shows up as an ingredient less than one percent of the time.

Table~\ref{tab:food_groups} categories all the ingredients into the 12 FAO food groups for calculating the HDDS. Not unexpectedly, given the prevalence of maize, cereals are the most frequently consumed food group (28\%). The next most common is vegetables (22\%) and spices, condiments, and beverages (19\%) because of the frequency with which salt is recorded. The least commonly consummed food groups are eggs (0.35\%), sweets (0.20\%), and fruits (0.16\%), which each make up less than one percent of consumption.

Finally, Table~\ref{tab:species} categories all the ingredients into scientific species. Considering ingredients by species, there are 66,262 total ingredient species in the data and 63 unique species. These totals are less than total and unique ingredients because things like maize and maize flour or pumpkin flesh and leaves are from the same species. Additionally, minerals like salt are not from a species and so are excluded. Unsurprisingly, \emph{zea mays} (maize) is the most common species, making up 38\% of observations. Next most common are tomato (\emph{lycopersicon esculentum}) and cassava (\emph{manihot esculenta}) at 12\% and 10\%. While we try to assign all ingredients to a distinct species, the level of detail recorded in the diaries forces us to aggregate some ingredients into large species categories. When households ate fish they rarely recorded the fish species and so all fish are categorized as \emph{fish spp}. Similarly all beef was categorized as \emph{bos taurus linnaeus, 1758}.

\setcounter{table}{0}
\renewcommand{\thetable}{A\arabic{table}}
\setcounter{figure}{0}
\renewcommand{\thefigure}{A\arabic{figure}}

\newpage

\begin{figure}[!htbp]
	\begin{minipage}{\linewidth}		
		\caption{Daily Food Diary, Days 1-6} \label{fig:diary1}
		\begin{center}
	       \includegraphics[width=.75\textwidth,keepaspectratio]{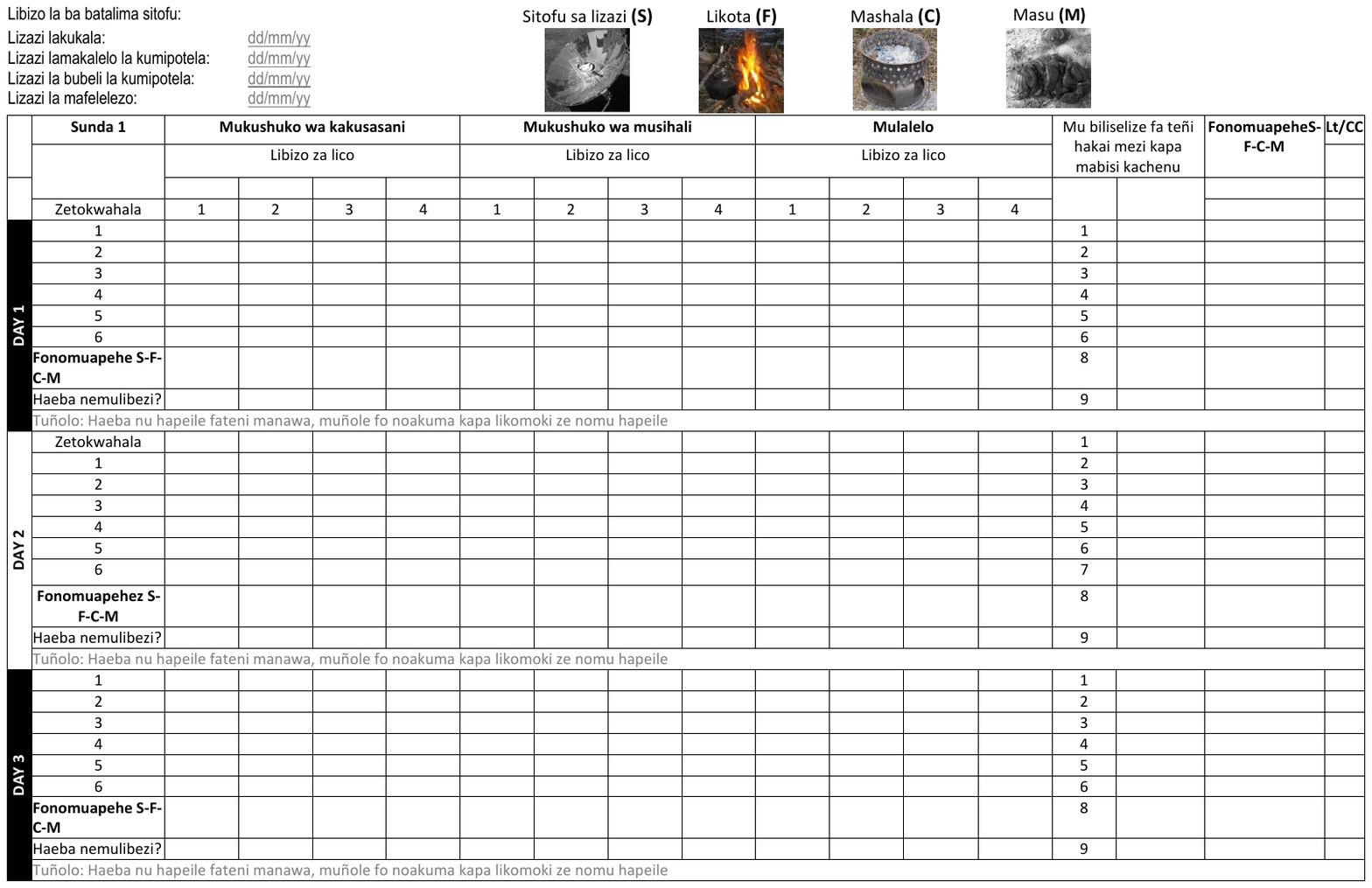}
	       \includegraphics[width=.75\textwidth,keepaspectratio]{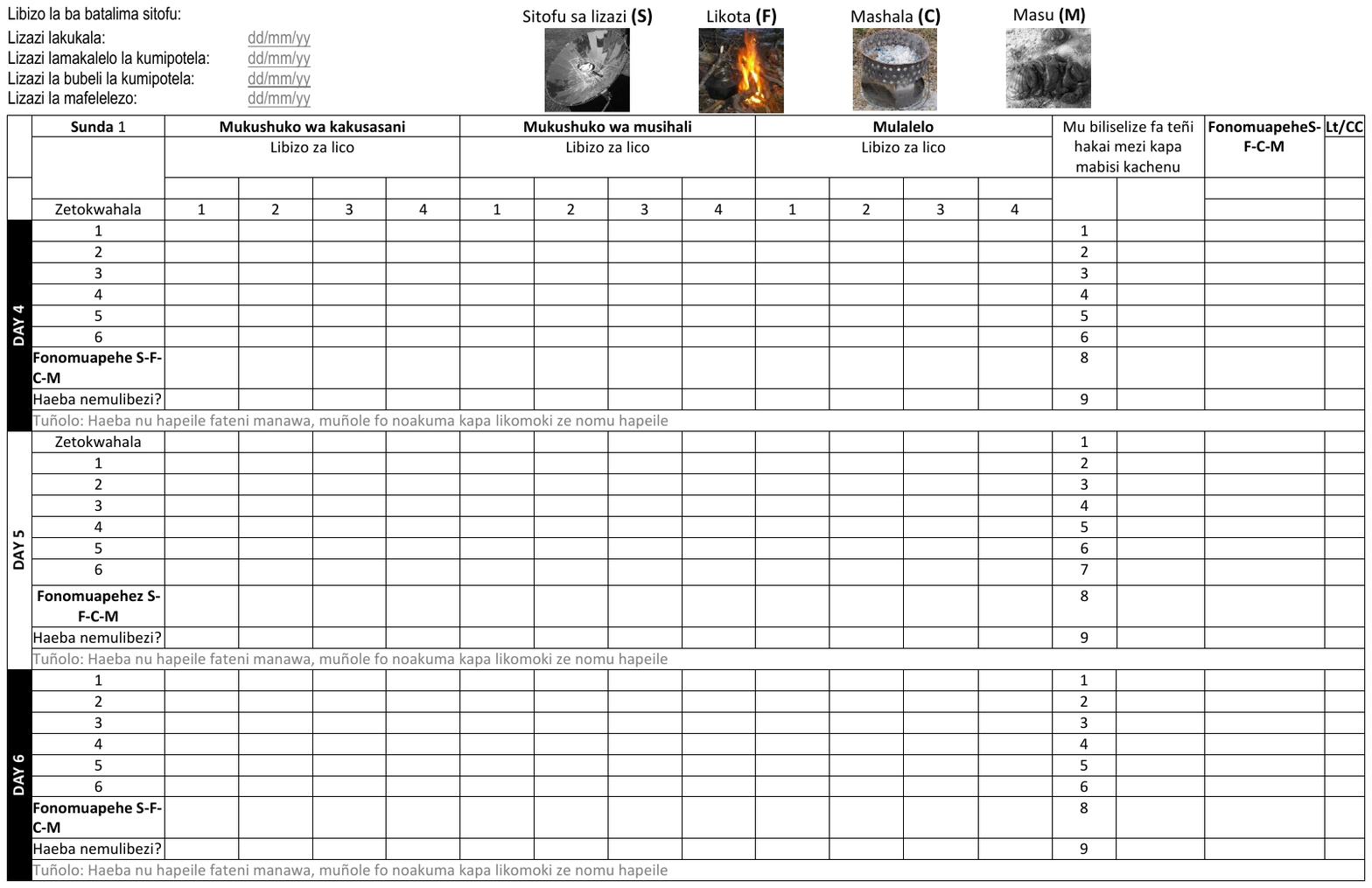}
		\end{center}
		\footnotesize  \textit{Note}: Participating households logged each ingredient in each dish for each meal (classified as breakfast, lunch, or dinner) throughout the six-week experiment. Participants also logged each dish's cooking method (solar stoves, firewood, charcoal, or dung, pictured at the top of the diary). Time and money spent on fuel collection and purchases were logged weekly.
	\end{minipage}	
\end{figure}

\begin{figure}[!htbp] 
	\begin{minipage}{\linewidth}	
	\caption{Daily Food Diary, Day 7, and Weekly Fuel Diary} 	\label{fig:diary3}
		\begin{center}
	\includegraphics[width=.75\textwidth,keepaspectratio]{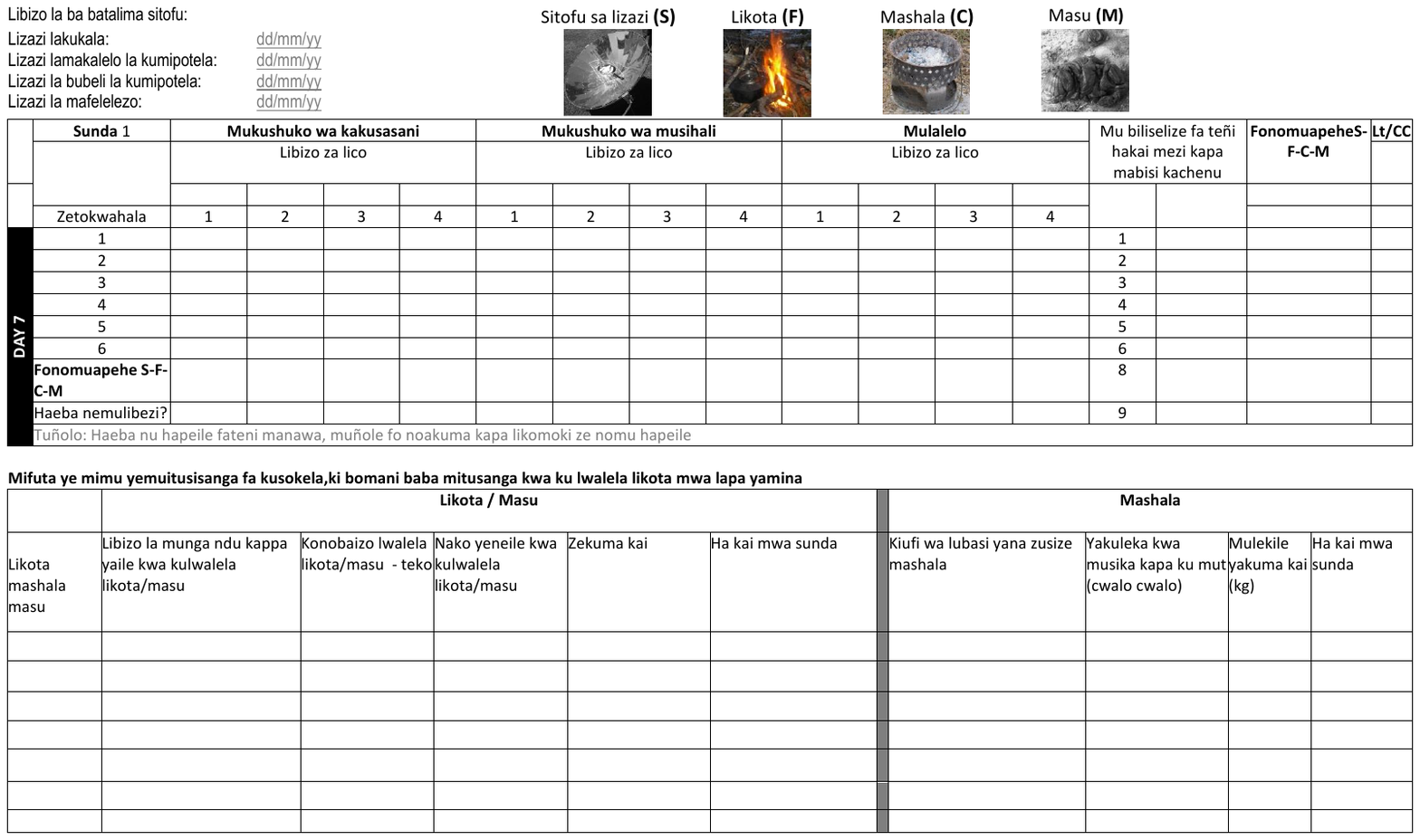}
		\end{center}
		\footnotesize  \textit{Note}: Participating households logged each ingredient in each dish for each meal (classified as breakfast, lunch, or dinner) throughout the six-week experiment. Participants also logged each dish's cooking method (solar stoves, firewood, charcoal, or dung, pictured at the top of the diary). Time and money spent on fuel collection and purchases were logged weekly.
	\end{minipage}
\end{figure}
 
\begin{landscape}
\begin{figure}[!htbp]
	\begin{minipage}{\linewidth}		
		\caption{Response Rates Over Time by Treatment} 	\label{fig:response}
		\begin{center}
	       \includegraphics[width=.49\textwidth,keepaspectratio]{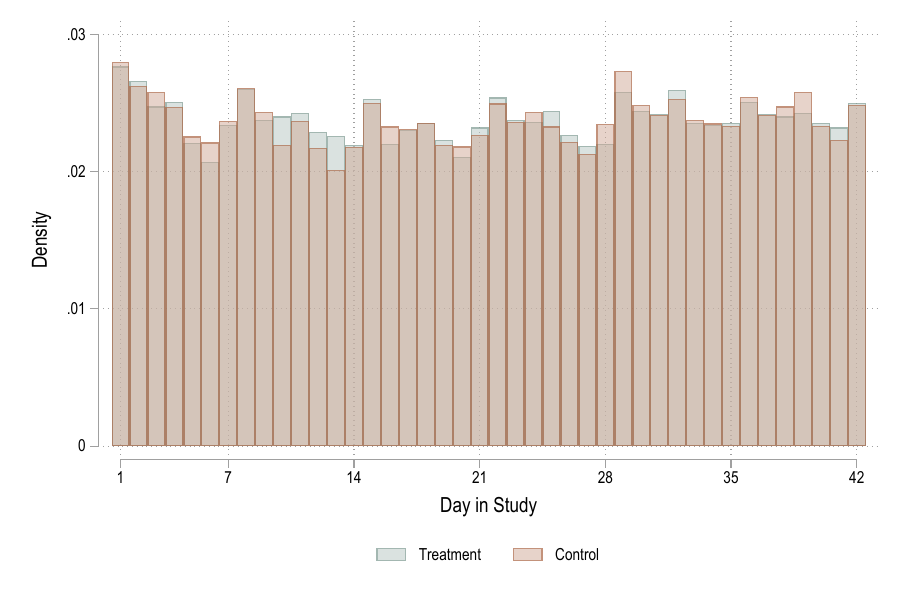}
	       \includegraphics[width=.49\textwidth,keepaspectratio]{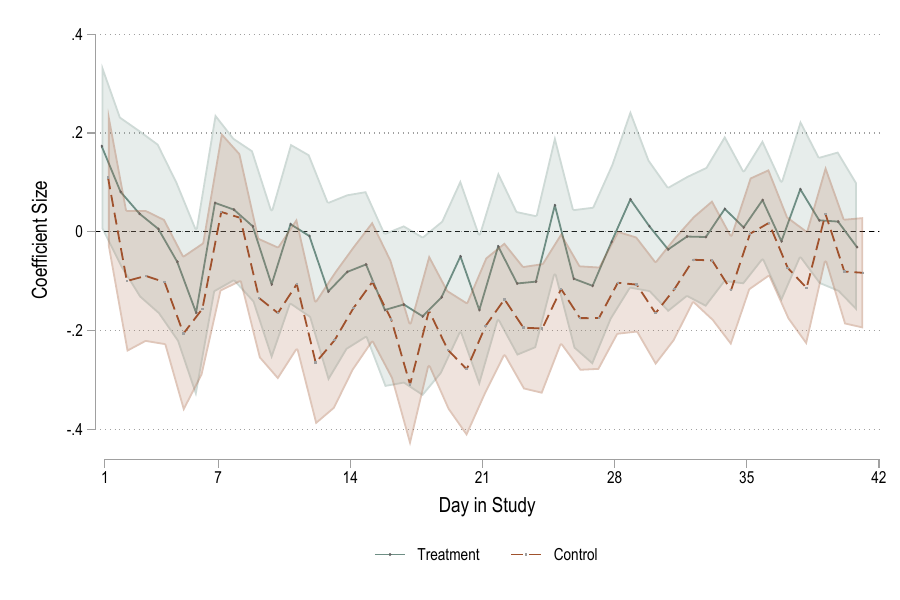}
		\end{center}
		\footnotesize  \textit{Note}: The left hand side panel of the figure presents the distribution of households recording information in their food diary on each day in the experiment, broken down by treatment and control. If every participant recorded information in their food diary on every day, the distribution would be uniform across time. The right hand side of the figure presents results from an event study style regressions in which the number of ingredient recorded in day is regressed on a set of dummies indicating each day in the experiment. Regressions are conducted by treatment status with standard errors clustered at the household. The omitted day is the last day in the experiment.
	\end{minipage}	
\end{figure}
\end{landscape}

\begin{table}[!htbp]	\centering
    \caption{List of Ingredients Recorded in Food Diaries\label{tab:ingredients}}
	\scalebox{1}
	{ \setlength{\linewidth}{.2cm}\newcommand{\contents}
		{\input{tables/ing_tab}}
	\setbox0=\hbox{\input{tables/fuel_late}}
    \setlength{\linewidth}{\wd0-2\tabcolsep-.25em}
    \input{tables/fuel_late}}
\end{table}

\begin{table}[!htbp]	\centering
    \caption{Ingredients Categorized by Food Group\label{tab:food_groups}}
	\scalebox{1}
	{ \setlength{\linewidth}{.2cm}\newcommand{\contents}
		{\input{tables/fg_tab}}
	\setbox0=\hbox{\input{tables/fuel_late}}
    \setlength{\linewidth}{\wd0-2\tabcolsep-.25em}
    \input{tables/fuel_late}}
\end{table}

\begin{table}[!htbp]	\centering
    \caption{Ingredients Categorized by Scientific Species\label{tab:species}}
	\scalebox{1}
	{ \setlength{\linewidth}{.2cm}\newcommand{\contents}
		{\input{tables/sci_tab}}
	\setbox0=\hbox{\input{tables/fuel_late}}
    \setlength{\linewidth}{\wd0-2\tabcolsep-.25em}
    \input{tables/fuel_late}}
\end{table}


\clearpage

\section{Summary Statistics and Cloud Cover Calculation} \label{app:summary}

In this appendix we report on variable creation and summary statistics for those variables used in the analysis. Table~\ref{tab:var_def} provides detailed variable definitions for treatment, outcomes, and controls.

Table~\ref{tab:out_tab} reports the mean and standard deviation of the main dietary outcome variables as well as the frequency of solar stove use in dish preparation. Solar stoves are used to cook 18\% of the 30,314 dishes prepared over the length of the experiment, meaning a solar stove was used 5,464 times. The standard recall period of calculating HDDS or species richness is 24 hours, so looking at mean values for those variables per day we see an HDDS of about six and a SR of about seven. That means in the average day a households consumes food from six of 12 food groups and from seven distinct species. Households eat legumes less than once a day (0.68) or about four times a week.

Table~\ref{tab:cook_tab} reports on cooking frequency. Households prepare about one dish for breakfast and two for lunch and dinner. In terms of meals skipped, recall that households recorded data for 42 days. In that time, households skipped breakfast an average of 10 times, which is nearly 25\% of all breakfasts. Lunch and dinner were skipped much less frequently (three and five times), which is around 10\% of each meal. Overall the 126 possible meals a household could eat in the 42 days, the average household skipped 20 meals.

Table~\ref{tab:fuel_tab} reports on fuel collection. Households, on average, spent 115 minutes collecting firewood. This amounts to around 70 kg of firewood to supplement the six kg of charcoal households purchased, on average. The market value of all fuel collected or purchased on a weekly basis is just under 10 USD or about 520 USD a year. For context, average annual income for a household in Zambia is around 19,900 USD adjusted for purchasing power parity. Thus, the average household spends three percent of their income on cooking fuel. Recall, the solar stoves cost 85 USD and can last up to 10 years.

Tables~\ref{tab:sum_con} and~\ref{tab:frq_cat} report summary statistics for the control variables used in some regressions. The average age of the head of household is 48 years and the average household size is seven. Seventy percent of households were headed by a woman, with half having at least a secondary education. Households were uneveningly distributed across AAS group and village, reflecting the stratification we did during the randomization process. Because of this stratification, we include AAS and village fixed effects in every regression.

For cloud cover, we use the Landsat Collection 1 Level-1 band Quality Assessment band (BQA) from imagery taken throughout the year 2016 (tile path 175; row 071). We reclassify the pixels on the BQA with high cloud or cloud shadow confidence attributes as 1/cloud cover to delimit the cloud area. Pixels with other attributes (e.g., low or medium confidence) were reclassified as 0/no cloud cover. We calculate the cloud area in a 5km radius around each village. Table~\ref{tab:cloud_cover} reports on cloud cover per village per week and Figure~\ref{fig:cloud_cover} reproduces the actual Landsat iamges. 

\setcounter{table}{0}
\renewcommand{\thetable}{B\arabic{table}}
\setcounter{figure}{0}
\renewcommand{\thefigure}{B\arabic{figure}}

\begin{table}[htbp]	\centering
	\caption{Definitions of Variables Used in Empirical Analysis} \label{tab:var_def}
	\scalebox{0.9}
	{ \setlength{\linewidth}{.1cm}\newcommand{\contents}
		{\begin{tabular}{ll}
			\\[-1.8ex]\hline 
			\hline \\[-1.8ex]
			\multicolumn{1}{l}{Variable} & \multicolumn{1}{l}{Description} \\  
                \midrule
			\multicolumn{2}{l}{\emph{Panel A: Treatments}} \\
			\multicolumn{1}{l}{Solar Stove Assignment} & \multicolumn{1}{p{11cm}}{$=1$ if household was randomly assigned a solar stove} \\
			\multicolumn{1}{l}{Solar Stove Use} & \multicolumn{1}{p{11cm}}{$=1$ if household used a solar stove to prepare a given dish} \\
			\midrule 
			\multicolumn{2}{l}{\emph{Panel B: Outcomes}} \\			
                \multicolumn{1}{l}{Household Dietary Diversity Score} & \multicolumn{1}{p{11cm}}{Count of unique food groups used in dish/meal/day/weel/overall (potential range: 0 - 12)}\\  
                \multicolumn{1}{l}{Dietary Species Richness} & \multicolumn{1}{p{11cm}}{Count of unique species used in dish/meal/day/weel/overall}\\       
                \multicolumn{1}{l}{Legume Consumption} & \multicolumn{1}{p{11cm}}{Count of the number of times legumes were eaten in dish/meal/day/weel/overall}\\        
                \multicolumn{1}{l}{Dishes per Meal} & \multicolumn{1}{p{11cm}}{Count of the number of dishes in breakfast/lunch/dinner/overall}\\             
                \multicolumn{1}{l}{Meals Skipped} & \multicolumn{1}{p{11cm}}{Count of the number of times breakfast/lunch/dinner was skipped or how many total meals were skipped}\\       
                \multicolumn{1}{l}{Firewood Time} & \multicolumn{1}{p{11cm}}{Amount of time in minutes that a household spent collecting firewood or cow dung in a week}\\        
                \multicolumn{1}{l}{Firewood Quantity} & \multicolumn{1}{p{11cm}}{Quantity in kg of firewood or cow dung a household collected or purchased in a week}\\        
                \multicolumn{1}{l}{Charcoal Quantity} & \multicolumn{1}{p{11cm}}{Quantity in kg of charcoal a hosuehold produced or purchased in a week}\\        
                \multicolumn{1}{l}{Fuel Value} & \multicolumn{1}{p{11cm}}{Total value in USD of all fuel used in a week valued at market prices}\\          
			\midrule 
			\multicolumn{2}{l}{\emph{Panel C: Controls}} \\
			\multicolumn{1}{l}{Age} & \multicolumn{1}{p{11cm}}{Age of household head in years} \\
                \multicolumn{1}{l}{Cloud Cover}& \multicolumn{1}{p{11cm}}{Percentage of cloud cover for the week} \\
                \multicolumn{1}{l}{Education}& \multicolumn{1}{p{11cm}}{Indicator for Education of household head (None, Primary, Secondary, or Higher)} \\
                \multicolumn{1}{l}{Gender} & \multicolumn{1}{p{11cm}}{$=1$ if gender of household head is female} \\
                \multicolumn{1}{l}{Household Size} & \multicolumn{1}{p{11cm}}{Total number of individuals in the household}\\
                \multicolumn{1}{l}{Tropical Livestock Index} & \multicolumn{1}{p{11cm}}{Tropical Livestock Unit based on FAO conversion factors} \\
                \multicolumn{1}{l}{Asset Index} & \multicolumn{1}{p{11cm}}{Index baed on principal component analysis of durable goods owned by household} \\
                \multicolumn{1}{l}{AAS group} & \multicolumn{1}{p{11cm}}{Indicator for each AAS group (none, learning plots, cooking clubs, both)} \\
                \multicolumn{1}{l}{Village} & \multicolumn{1}{p{11cm}}{Indicator for each village} \\
			\\[-1.8ex]\hline 
			\hline \\[-1.8ex]
			\multicolumn{2}{p{\linewidth}}{\footnotesize  \textit{Note}: The table presents definitions of variables used in the analysis.} \\
		\end{tabular}}
	\setbox0=\hbox{\input{tables/fuel_late}}
    \setlength{\linewidth}{\wd0-2\tabcolsep-.25em}
    \input{tables/fuel_late}}
\end{table}

\begin{table}[!htbp]	\centering
    \caption{Summary Statistics for Intermediate and Dietary Outcomes\label{tab:out_tab}}
	\scalebox{1}
	{ \setlength{\linewidth}{.2cm}\newcommand{\contents}
		{\input{tables/out_tab}}
	\setbox0=\hbox{\input{tables/fuel_late}}
    \setlength{\linewidth}{\wd0-2\tabcolsep-.25em}
    \input{tables/fuel_late}}
\end{table}

\begin{table}[!htbp]	\centering
    \caption{Summary Statistics for Cooking Frequency\label{tab:cook_tab}}
	\scalebox{1}
	{ \setlength{\linewidth}{.2cm}\newcommand{\contents}
		{\input{tables/cook_tab}}
	\setbox0=\hbox{\input{tables/fuel_late}}
    \setlength{\linewidth}{\wd0-2\tabcolsep-.25em}
    \input{tables/fuel_late}}
\end{table}

\begin{table}[!htbp]	\centering
    \caption{Summary Statistics for Fuel Collection\label{tab:fuel_tab}}
	\scalebox{1}
	{ \setlength{\linewidth}{.2cm}\newcommand{\contents}
		{\input{tables/fuel_tab}}
	\setbox0=\hbox{\input{tables/fuel_late}}
    \setlength{\linewidth}{\wd0-2\tabcolsep-.25em}
    \input{tables/fuel_late}}
\end{table}

\begin{table}[!htbp]	\centering
    \caption{Summary Statistics for Continuous Control Variables\label{tab:sum_con}}
	\scalebox{1}
	{ \setlength{\linewidth}{.2cm}\newcommand{\contents}
		{\input{tables/convars_tab}}
	\setbox0=\hbox{\input{tables/fuel_late}}
    \setlength{\linewidth}{\wd0-2\tabcolsep-.25em}
    \input{tables/fuel_late}}
\end{table}

\begin{table}[!htbp]	\centering
    \caption{Categorical Covariate Frequency Table\label{tab:frq_cat}}
	\scalebox{1}
	{ \setlength{\linewidth}{.2cm}\newcommand{\contents}
		{\input{tables/catvars_tab}}
	\setbox0=\hbox{\input{tables/fuel_late}}
    \setlength{\linewidth}{\wd0-2\tabcolsep-.25em}
    \input{tables/fuel_late}}
\end{table}

\begin{table}[!htbp] \centering 
	\caption{Percentage of village area with clouds or cloud shadows} \label{tab:cloud_cover}
	\scalebox{1}
	{\setlength{\linewidth}{.1cm}\newcommand{\contents}
	{\begin{tabular}{ccccccc} 
		\hline 
		\hline \\[-1.8ex] 
		\multicolumn{3}{c}{Data Aquired} & \multicolumn{3}{c}{Villages} & \multicolumn{1}{c}{Project}  \\ 
		\multicolumn{1}{c}{Month} & \multicolumn{1}{c}{Day} & \multicolumn{1}{c}{Year} & \multicolumn{1}{c}{Village L} & \multicolumn{1}{c}{Village M} & \multicolumn{1}{c}{Village N} & \multicolumn{1}{c}{Ongoing} \\ 
		\hline \\[-1.8ex] 
		1       & 25   & 2016   & 58.0   & 91.5   & 25.2   & No \\
		2       & 10   & 2016   & 13.6   & 5.69   & 28.8   & No \\
		2       & 26   & 2016   & 100    & 100    & 100    & Yes \\
		3       & 13   & 2016   & 11.8   & 1.70   & 35.8   & Yes \\
		3       & 29   & 2016   & 100    & 2.06   & 0.00   & Yes \\
		4       & 14   & 2016   & 0.00   & 0.00   & 0.00   & Yes \\
		4       & 30   & 2016   & 0.00   & 0.01   & 0.00   & Yes \\
		5       & 16   & 2016   & 0.00   & 0.00   & 0.00   & No \\
		6       & 1    & 2016   & 0.00   & 0.00   & 0.00   & No \\
		6       & 17   & 2016   & 0.00   & 0.00   & 0.00   & No \\
		7       & 3	   & 2016   & 0.00   & 0.00   & 0.00   & No \\
		7       & 19   & 2016   & 0.00   & 0.00   & 0.00   & No \\
		8       & 4	   & 2016   & 0.00   & 0.00   & 0.00   & No \\
		8       & 20   & 2016   & 0.00   & 0.00   & 0.00   & No \\
		9       & 5	   & 2016   & 0.10   & 0.00   & 0.00   & No \\
		9       & 21   & 2016   & 0.00   & 0.12   & 0.00   & No \\
		10      & 7	   & 2016   & 0.00   & 0.00   & 0.00   & No \\
		10      & 23   & 2016   & 0.00   & 0.00   & 33.5   & No \\
		11      & 8	   & 2016   & 0.00   & 0.00   & 0.00   & No \\
		11      & 24   & 2016   & 99.9   & 54.1   & 1.46   & No \\
		12      & 10   & 2016   & 53.8   & 91.0   & 0.88   & No \\
		12      & 26   & 2016   & 26.9   & 100    & 23.8   & No \\
		\hline 
		\hline \\[-1.8ex]
		\multicolumn{7}{p{\linewidth}}{\footnotesize \textit{Note}: Percentage of the area in a buffer of 5km around each village with clouds or clouds shadows. See Figure 1 for a visual identification of the cloud cover area throughout the year 2016.}
	\end{tabular}}
		\setbox0=\hbox{\input{tables/fuel_late}}
		\setlength{\linewidth}{\wd0-2\tabcolsep-.25em}
		\input{tables/fuel_late}}
\end{table}

\begin{figure}[!htbp] \centering
	\caption{Cloud cover during experiment} 	\label{fig:cloud_cover}
	\begin{minipage}{\textwidth}
		\includegraphics[width=.95\textwidth,keepaspectratio]{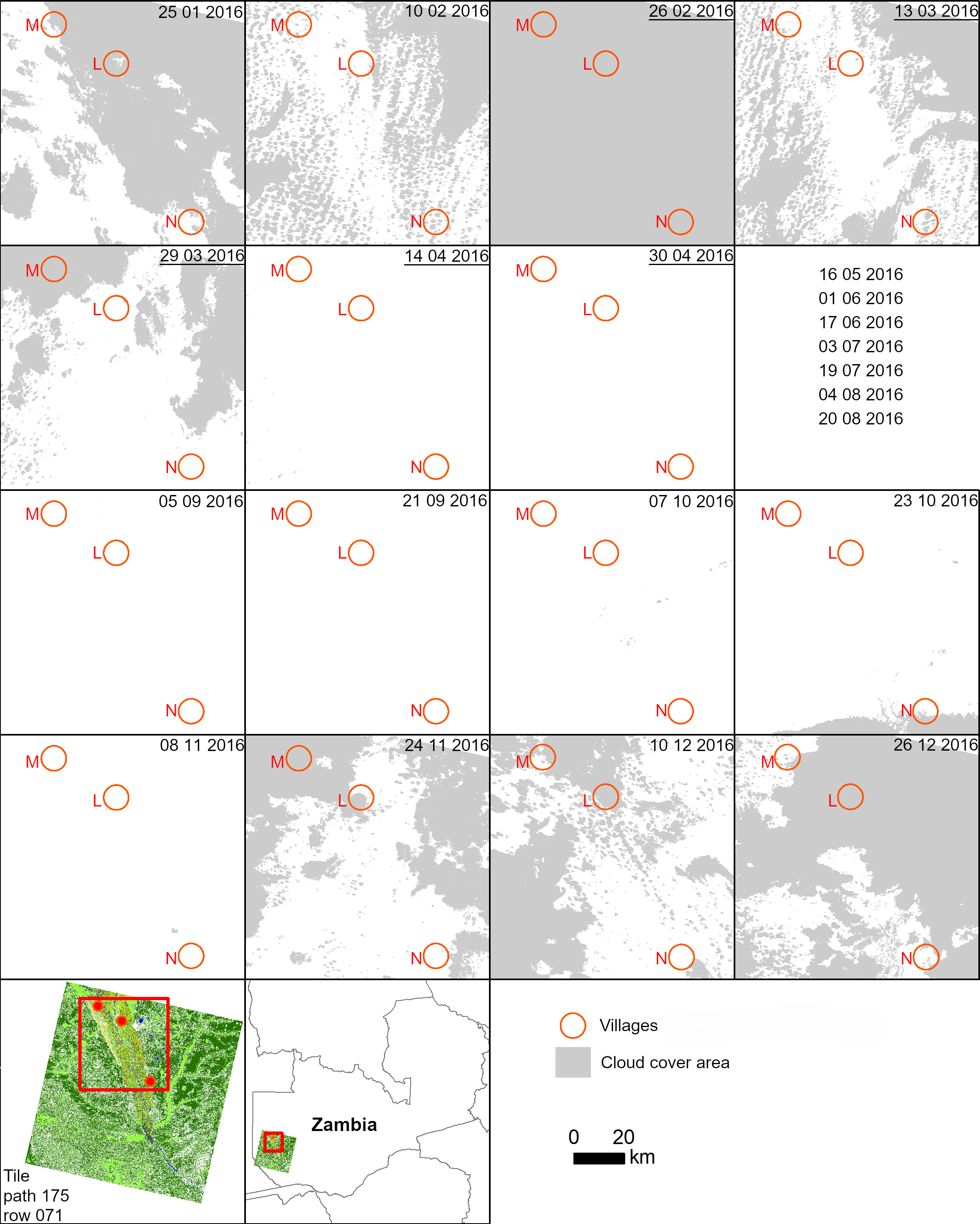} \\
		\footnotesize  \textit{Note}: Cloud cover area delimited with Landsat 8 Enhanced Thematic Mapper ™ path 175 row 071 in a radius of 5km around each village where the solar project took place during the end of February until the end of April 2016. The whole tile area had a cloud cover of 0\% from 16 05 2016 until 20 08 2016.
	\end{minipage}
\end{figure}


\clearpage
\section{Robustness Checks}  \label{app:robust}

In this appendix we present results of robustness checks on our main outcomes of interest. As a first check, we calculate the local average treatment effect (LATE) as opposed to the intent-to-treat (ITT) effects reported in the paper. Second, we construct our outcomes of interest as averages instead of simply counts. Both were pre-specified in our pre-analysis plan.

We are able to calculate the LATE because we know who was assigned to treatment and who used a stove and when. Assignment to treatment is random while use of a stove is not random. Our primary results that rely on random assignment are thus ITTs, since we do not control who uses a stove. This is the most straightfoward approach to estimating a treatment effect though it will be an underestimates of the average treatment effect (ATE) because households do not use the stove for every dish or at every meal. We could only estimate the ATE (the effect of using a stove on outcomes) if we dictated, at random, when a household used a stove. To determine if the null effects of solar stoves on dietary composition and cooking frequency are due to this underestmate, we calculate the LATE. We regress our outcomes of interest on whether or not a household used a solar stove for a specific dish. At levels of aggregations greater than a dish (meal, day, week, etc.) we use the share of dishes that were cooked with a solar stove. Since the choice to use a solar stove to cook a specific dish is not random, we instrument for the use of the solar stove with the random assignment to treatment.

As can be seen in Table~\ref{tab:late_diverse}, coefficients on our measures of dietary diversity remain stastically insignificant. Though effect sizes are larger, as we would expect when comparing the LATE to the ITT, the estimates are less precise (larger standard errors), as we would also expect when using an instrumental variable. A similar pattern is present in Table~\ref{tab:late_numdish}. While the LATE produces larger coefficients, they all remain insignificant as the standard errors also increase in size. Finally, Table~\ref{tab:fuel_late} we report LATE estimates of the use of solar stoves on fuel. Again, coefficients increase in size. The more frequently a household uses a stove, the large time and cost savings that household experiences. While coefficients on firewood quantity and overall value remain significant, savings in terms of time becomes not significant due to the inflation of standard errors.

Overall, our ITT results are robust to substitution with the LATE. While the ITT and LATE measure treatment effects in different populations, the two are related in well known ways. The LATE tends to have larger coefficients than the ITT since the ITT includes non-compliers. Additionally, since the LATE is an IV approach to estimating the treatment effect, it tends to be a less efficient estimator than the ITT estimated via OLS. In our results, we see these same relations hold, which gives us confidence in the reliability of our ITT estimates presented in the paper. 


\setcounter{table}{0}
\renewcommand{\thetable}{C\arabic{table}}
\setcounter{figure}{0}
\renewcommand{\thefigure}{C\arabic{figure}}

\begin{landscape}
\begin{table}[!htbp]	\centering
    \caption{LATE Estimates of Solar Stove Use on Dietary Composition} \label{tab:late_diverse}
	\scalebox{1}
	{ \setlength{\linewidth}{.2cm}\newcommand{\contents}
		{\input{tables/late_out.tex}}
	\setbox0=\hbox{\input{tables/fuel_late}}
    \setlength{\linewidth}{\wd0-2\tabcolsep-.25em}
    \input{tables/fuel_late}}
\end{table}
\end{landscape}

\begin{landscape}
\begin{table}[!htbp]	\centering
    \caption{LATE Estimates of Solar Stove Use on Frequency of Cooking} \label{tab:late_numdish}
	\scalebox{1}
	{ \setlength{\linewidth}{.2cm}\newcommand{\contents}
		{\input{tables/late_numdish}}
	\setbox0=\hbox{\input{tables/fuel_late}}
    \setlength{\linewidth}{\wd0-2\tabcolsep-.25em}
    \input{tables/fuel_late}}
\end{table}
\end{landscape}

\begin{landscape}
\begin{table}[!htbp]	\centering
    \caption{LATE Estimates of Solar Stove Use on Fuel Collection} \label{tab:fuel_late}
	\scalebox{1}
	{ \setlength{\linewidth}{.2cm}\newcommand{\contents}
		{\input{tables/fuel_late}}
	\setbox0=\hbox{\input{tables/fuel_late}}
    \setlength{\linewidth}{\wd0-2\tabcolsep-.25em}
    \input{tables/fuel_late}}
\end{table}
\end{landscape}

\end{document}

%% file: tables/ss_use.tex
\begin{tabular}{l*{10}{c}} \\[-1.8ex]\hline \hline \\[-1.8ex] 
& \multicolumn{2}{c}{Dish} & \multicolumn{2}{c}{Meal} & \multicolumn{2}{c}{Day} 
& \multicolumn{2}{c}{Week} & \multicolumn{2}{c}{Overall} \\ \cline{2-3} 
\cline{4-5} \cline{6-7} \cline{8-9} \cline{10-11} \\[-1.8ex] 
& \multicolumn{1}{c}{(1)} & \multicolumn{1}{c}{(2)} & \multicolumn{1}{c}{(3)} 
& \multicolumn{1}{c}{(4)} &\multicolumn{1}{c}{(5)} & \multicolumn{1}{c}{(6)} 
& \multicolumn{1}{c}{(7)} & \multicolumn{1}{c}{(8)} & \multicolumn{1}{c}{(9)} 
& \multicolumn{1}{c}{(10)} \\ \midrule
\midrule
Solar Stove         &       0.426\sym{***}&       0.424\sym{***}&       0.421\sym{***}&       0.421\sym{***}&       0.433\sym{***}&       0.431\sym{***}&       0.441\sym{***}&       0.440\sym{***}&       0.437\sym{***}&       0.436\sym{***}\\
                    &     (0.027)         &     (0.028)         &     (0.028)         &     (0.028)         &     (0.029)         &     (0.030)         &     (0.030)         &     (0.031)         &     (0.030)         &     (0.032)         \\
\midrule
Observations        &      30,314         &      29,707         &      15,896         &      15,578         &       6,013         &       5,891         &         912         &         894         &         156         &         153         \\
Covariates          &          No         &         Yes         &          No         &         Yes         &          No         &         Yes         &          No         &         Yes         &          No         &         Yes         \\
Adjusted R$^2$      &       0.281         &       0.292         &       0.319         &       0.334         &       0.483         &       0.497         &       0.591         &       0.608         &       0.670         &       0.671         \\
\hline \hline \\[-1.8ex] \multicolumn{11}{J{\linewidth}}{\small 
\noindent \textit{Note}: Dependent variable is the number of dishes, or 
the share of dishes in a gven meal, day, week, etc., for which a 
solar stove was used. All regressions include two levels of strata 
fixed effects: village and Agricultural and Aquatic Systems (AAS) group. 
For regressions with more than one observation per houhold (columns 1-8), 
we calculate Liang-Zeger cluster-robust standard errors since the unit 
of randomization is the household. For regressions with only one 
observation per household (columns 9-10), we calculate Eicker-Huber-White 
(EHW) robust standard errors. Standard errors are presented in 
parentheses (*** p$<$0.001, ** p$<$0.01, * p$<$0.05).}  \end{tabular}

%% file: tables/fuel_late.tex
\begin{tabular}{l*{8}{c}} \\[-1.8ex]\hline \hline \\[-1.8ex] 
& \multicolumn{2}{c}{Firewood} & \multicolumn{2}{c}{Firewood} 
& \multicolumn{2}{c}{Charcoal} & \multicolumn{2}{c}{Fuel} \\ 
& \multicolumn{2}{c}{Time (min)} & \multicolumn{2}{c}{Quantity (kg)} 
& \multicolumn{2}{c}{Quantity (kg)} & \multicolumn{2}{c}{Value (USD)} \\ 
\cline{2-3} \cline{4-5} \cline{6-7} \cline{8-9} \\[-1.8ex] 
& \multicolumn{1}{c}{(1)} & \multicolumn{1}{c}{(2)} & \multicolumn{1}{c}{(3)} 
& \multicolumn{1}{c}{(4)} &\multicolumn{1}{c}{(5)} & \multicolumn{1}{c}{(6)} 
& \multicolumn{1}{c}{(7)} & \multicolumn{1}{c}{(8)}  \\ \midrule 
\multicolumn{9}{l}{\emph{Panel A: Weekly Fuel Outcomes}} \\ 
\midrule
Solar Stove Use     &      -78.88         &      -80.73         &      -65.35\sym{*}  &      -71.81\sym{*}  &       -5.15         &       -6.93         &       -9.11\sym{*}  &      -10.17\sym{*}  \\
                    &     (42.11)         &     (46.32)         &     (29.32)         &     (35.15)         &      (4.74)         &      (5.23)         &      (3.75)         &      (4.51)         \\
\midrule
Mean in Control     &     127.1         &     127.1         &      79.02         &      79.02         &       7.639         &       7.639         &      11.19         &      11.19         \\
Observations        &         852         &         834         &         852         &         834         &         852         &         834         &         852         &         834         \\
Covariates          &          No         &         Yes         &          No         &         Yes         &          No         &         Yes         &          No         &         Yes         \\
Adjusted R$^2$      &       0.019         &       0.038         &       0.088         &       0.123         &       0.121         &       0.216         &       0.063         &       0.091         \\
\midrule \multicolumn{9}{l}{\emph{Panel B: Overall Fuel Outcomes}} \\ 
\midrule
Solar Stove Use     &     -454.2         &     -436.3         &     -343.8\sym{*}  &     -358.1\sym{*}  &      -35.74         &      -38.92         &      -48.99\sym{*}  &      -51.22\sym{*}  \\
                    &    (236.9)         &    (239.1)         &    (152.8)         &    (178.0)         &     (26.81)         &     (27.57)         &     (19.63)         &     (22.88)         \\
\midrule
Mean in Control     &     704.0         &     704.0         &     437.5         &     437.5         &      42.30         &      42.30         &      61.95         &      61.95         \\
Observations        &         156         &         153         &         156         &         153         &         156         &         153         &         156         &         153         \\
Covariates          &          No         &         Yes         &          No         &         Yes         &          No         &         Yes         &          No         &         Yes         \\
Adjusted R$^2$      &       0.008         &       0.010         &       0.107         &       0.115         &       0.120         &       0.250         &       0.074         &       0.077         \\
\hline \hline \\[-1.8ex] \multicolumn{9}{J{\linewidth}}{\small 
\noindent \textit{Note}: Dependent variables are different measure of 
 fuel collection at different levels of aggregation. 
In Panel A, we use values measured each week 
In Panel B, we sum weekly values to the overall six week total. 
All regressions include two levels of strata 
fixed effects: village and Agricultural and Aquatic Systems (AAS) group. 
Eicker-Huber-White (EHW) robust standard errors. Standard errors are presented in 
parentheses (*** p$<$0.001, ** p$<$0.01, * p$<$0.05).}  \end{tabular}

%% file: tables/diverse_out.tex
\begin{tabular}{l*{10}{c}} \\[-1.8ex]\hline \hline \\[-1.8ex] 
& \multicolumn{2}{c}{Dish} & \multicolumn{2}{c}{Meal} & \multicolumn{2}{c}{Day} 
& \multicolumn{2}{c}{Week} & \multicolumn{2}{c}{Overall} \\ \cline{2-3} 
\cline{4-5} \cline{6-7} \cline{8-9} \cline{10-11} \\[-1.8ex] 
& \multicolumn{1}{c}{(1)} & \multicolumn{1}{c}{(2)} & \multicolumn{1}{c}{(3)} 
& \multicolumn{1}{c}{(4)} &\multicolumn{1}{c}{(5)} & \multicolumn{1}{c}{(6)} 
& \multicolumn{1}{c}{(7)} & \multicolumn{1}{c}{(8)} & \multicolumn{1}{c}{(9)} 
& \multicolumn{1}{c}{(10)} \\ \midrule 
\multicolumn{11}{l}{\emph{Panel A: Household Dietary Diversity Score}} \\ 
\midrule
Solar Stove         &      -0.063         &      -0.058         &      -0.117         &      -0.114         &      -0.082         &      -0.129         &      -0.098         &      -0.142         &      -0.063         &      -0.109         \\
                    &     (0.062)         &     (0.059)         &     (0.110)         &     (0.110)         &     (0.147)         &     (0.148)         &     (0.175)         &     (0.182)         &     (0.185)         &     (0.196)         \\
\midrule
Mean in Control     &       2.332         &       2.332         &       3.908         &       3.908         &       5.668         &       5.668         &       8.168         &       8.168         &       9.768         &       9.768         \\
Observations        &      30,314         &      29,707         &      15,896         &      15,578         &       6,013         &       5,891         &         912         &         894         &         156         &         153         \\
Covariates          &          No         &         Yes         &          No         &         Yes         &          No         &         Yes         &          No         &         Yes         &          No         &         Yes         \\
Adjusted R$^2$      &       0.024         &       0.035         &       0.038         &       0.048         &       0.124         &       0.133         &       0.136         &       0.140         &       0.157         &       0.130         \\
\midrule \multicolumn{11}{l}{\emph{Panel B: Dietary Species Richness}} \\ 
\midrule
Solar Stove         &      -0.066         &      -0.077         &      -0.079         &      -0.102         &       0.029         &      -0.063         &      -0.161         &      -0.306         &      -0.321         &      -0.516         \\
                    &     (0.069)         &     (0.068)         &     (0.137)         &     (0.136)         &     (0.246)         &     (0.243)         &     (0.517)         &     (0.529)         &     (0.810)         &     (0.861)         \\
\midrule
Mean in Control     &       2.329         &       2.329         &       3.984         &       3.984         &       6.552         &       6.552         &      13.28         &      13.28         &      20.14         &      20.14         \\
Observations        &      30,314         &      29,707         &      15,896         &      15,578         &       6,013         &       5,891         &         912         &         894         &         156         &         153         \\
Covariates          &          No         &         Yes         &          No         &         Yes         &          No         &         Yes         &          No         &         Yes         &          No         &         Yes         \\
Adjusted R$^2$      &       0.028         &       0.033         &       0.052         &       0.062         &       0.148         &       0.160         &       0.133         &       0.146         &       0.076         &       0.074         \\
\midrule \multicolumn{11}{l}{\emph{Panel C: Count of Legume Consumption}} \\ 
\midrule
Solar Stove         &       0.010         &       0.007         &       0.021         &       0.015         &       0.067         &       0.045         &       0.438         &       0.283         &       2.520         &       1.574         \\
                    &     (0.013)         &     (0.013)         &     (0.026)         &     (0.026)         &     (0.072)         &     (0.070)         &     (0.493)         &     (0.483)         &     (2.996)         &     (3.024)         \\
\midrule
Mean in Control     &       0.126         &       0.126         &       0.238         &       0.238         &       0.620         &       0.620         &       4.068         &       4.068         &      23.77         &      23.77         \\
Observations        &      30,314         &      29,707         &      15,896         &      15,578         &       6,013         &       5,891         &         912         &         894         &         156         &         153         \\
Covariates          &          No         &         Yes         &          No         &         Yes         &          No         &         Yes         &          No         &         Yes         &          No         &         Yes         \\
Adjusted R$^2$      &       0.006         &       0.009         &       0.018         &       0.022         &       0.048         &       0.057         &       0.112         &       0.132         &       0.137         &       0.133         \\
\hline \hline \\[-1.8ex] \multicolumn{11}{J{\linewidth}}{\small 
\noindent \textit{Note}: Dependent variables are different measure of 
 household dietary composition. In Panel A, we use dietary diversity 
score. In Panel B, we use species richness. In Panel C, we calculate 
the number of times legumes are eaten. All regressions include two levels of strata 
fixed effects: village and Agricultural and Aquatic Systems (AAS) group. 
For regressions with more than one observation per houhold (columns 1-8), 
we calculate Liang-Zeger cluster-robust standard errors since the unit 
of randomization is the household. For regressions with only one 
observation per household (columns 9-10), we calculate Eicker-Huber-White 
(EHW) robust standard errors. Standard errors are presented in 
parentheses (*** p$<$0.001, ** p$<$0.01, * p$<$0.05).}  \end{tabular}

%% file: tables/freq_out.tex
\begin{tabular}{l*{6}{c}} \\[-1.8ex]\hline \hline \\[-1.8ex] 
& \multicolumn{2}{c}{Day} & \multicolumn{2}{c}{Week} & 
\multicolumn{2}{c}{Overall} \\ \cline{2-3} 
\cline{4-5} \cline{6-7} \\[-1.8ex] 
& \multicolumn{1}{c}{(1)} & \multicolumn{1}{c}{(2)} & \multicolumn{1}{c}{(3)} 
& \multicolumn{1}{c}{(4)} &\multicolumn{1}{c}{(5)} & \multicolumn{1}{c}{(6)} 
\\ \midrule 
\multicolumn{7}{l}{\emph{Panel A: Number of Dishes Prepared}} \\ 
\midrule
Solar Stove         &       0.164         &       0.107         &       1.211         &       0.664         &       7.175         &       3.044         \\
                    &     (0.151)         &     (0.148)         &     (1.381)         &     (1.385)         &     (9.527)         &     (9.730)         \\
\midrule
Mean in Control     &       4.920         &       4.920         &      32.27         &      32.27         &     188.5         &     188.5         \\
Observations        &       6,013         &       5,891         &         912         &         894         &         156         &         153         \\
Covariates          &          No         &         Yes         &          No         &         Yes         &          No         &         Yes         \\
Adjusted R$^2$      &       0.135         &       0.152         &       0.144         &       0.161         &       0.118         &       0.133         \\
\midrule \multicolumn{7}{l}{\emph{Panel B: Number of Meals Skipped}} \\ 
\midrule
Solar Stove         &      -0.052         &      -0.029         &      -0.419         &      -0.191         &      -2.591         &      -0.770         \\
                    &     (0.044)         &     (0.044)         &     (0.528)         &     (0.535)         &     (3.910)         &     (3.993)         \\
\midrule
Mean in Control     &       0.396         &       0.396         &       3.924         &       3.924         &      26.24         &      26.24         \\
Observations        &       6,013         &       5,891         &         912         &         894         &         156         &         153         \\
Covariates          &          No         &         Yes         &          No         &         Yes         &          No         &         Yes         \\
Adjusted R$^2$      &       0.048         &       0.067         &       0.056         &       0.078         &       0.020         &       0.091         \\
\hline \hline \\[-1.8ex] \multicolumn{7}{J{\linewidth}}{\small 
\noindent \textit{Note}: Dependent variables are different measure of 
 frequency of cooking. In Panel A, we use the number of dishes cooked 
in a meal. In Panel B, we use the number of meals skipped. 
All regressions include two levels of strata 
fixed effects: village and Agricultural and Aquatic Systems (AAS) group. 
Eicker-Huber-White (EHW) robust standard errors. Standard errors are presented in 
parentheses (*** p$<$0.001, ** p$<$0.01, * p$<$0.05).}  \end{tabular}

%% file: tables/fuel_out.tex
\begin{tabular}{l*{8}{c}} \\[-1.8ex]\hline \hline \\[-1.8ex] 
& \multicolumn{2}{c}{Firewood} & \multicolumn{2}{c}{Firewood} 
& \multicolumn{2}{c}{Charcoal} & \multicolumn{2}{c}{Fuel} \\ 
& \multicolumn{2}{c}{Time (min)} & \multicolumn{2}{c}{Quantity (kg)} 
& \multicolumn{2}{c}{Quantity (kg)} & \multicolumn{2}{c}{Value (USD)} \\ 
\cline{2-3} \cline{4-5} \cline{6-7} \cline{8-9} \\[-1.8ex] 
& \multicolumn{1}{c}{(1)} & \multicolumn{1}{c}{(2)} & \multicolumn{1}{c}{(3)} 
& \multicolumn{1}{c}{(4)} &\multicolumn{1}{c}{(5)} & \multicolumn{1}{c}{(6)} 
& \multicolumn{1}{c}{(7)} & \multicolumn{1}{c}{(8)}  \\ \midrule 
\multicolumn{9}{l}{\emph{Panel A: Weekly Fuel Outcomes}} \\ 
\midrule
Solar Stove         &      -43.72\sym{*}  &      -44.65\sym{*}  &      -26.91\sym{*}  &      -28.99\sym{*}  &       -2.01         &       -2.71         &       -3.74\sym{*}  &       -4.09\sym{*}  \\
                    &     (19.75)         &     (21.25)         &     (11.67)         &     (13.75)         &      (2.00)         &      (2.20)         &      (1.49)         &      (1.76)         \\
\midrule
Mean in Control     &     131.5         &     131.5         &      78.44         &      78.44         &       7.455         &       7.455         &      11.092         &      11.092         \\
Observations        &         870         &         852         &         870         &         852         &         870         &         852         &         870         &         852         \\
Covariates          &          No         &         Yes         &          No         &         Yes         &          No         &         Yes         &          No         &         Yes         \\
Adjusted R$^2$      &       0.024         &       0.047         &       0.112         &       0.152         &       0.116         &       0.209         &       0.088         &       0.121         \\
\midrule \multicolumn{9}{l}{\emph{Panel B: Overall Fuel Outcomes}} \\ 
\midrule
Solar Stove         &     -251.0\sym{*}  &     -244.0\sym{*}  &     -147.4\sym{*}  &     -152.6\sym{*}  &      -14.77         &      -16.11         &      -20.94\sym{*}  &      -21.77\sym{*}  \\
                    &    (115.6)         &    (119.6)         &     (65.37)         &     (76.99)         &     (11.99)         &     (12.65)         &      (8.392)         &      (9.904)         \\
\midrule
Mean in Control     &     738.2         &     738.2         &     440.4         &     440.4         &      41.85         &      41.85         &      62.28         &      62.28         \\
Observations        &         157         &         154         &         157         &         154         &         157         &         154         &         157         &         154         \\
Covariates          &          No         &         Yes         &          No         &         Yes         &          No         &         Yes         &          No         &         Yes         \\
Adjusted R$^2$      &       0.019         &       0.023         &       0.130         &       0.144         &       0.113         &       0.241         &       0.098         &       0.103         \\
\hline \hline \\[-1.8ex] \multicolumn{9}{J{\linewidth}}{\small 
\noindent \textit{Note}: Dependent variables are different measure of 
 fuel collection at different levels of aggregation. 
In Panel A, we use values measured each week 
In Panel B, we sum weekly values to the overall six week total. 
All regressions include two levels of strata 
fixed effects: village and Agricultural and Aquatic Systems (AAS) group. 
Eicker-Huber-White (EHW) robust standard errors. Standard errors are presented in 
parentheses (*** p$<$0.001, ** p$<$0.01, * p$<$0.05).}  \end{tabular}

%% file: tables/ing_tab.tex
\begin{tabular}{l*{2}{c}} \\ [-1.8ex]\hline \hline \\[-1.8ex] 
            &   Frequency&     Percent\\
\midrule
salt        &      13,669&       14.60\\
porridge    &      10,522&       11.24\\
oil         &       9,844&       10.52\\
maize       &       8,262&        8.83\\
tomato      &       7,977&        8.52\\
maize flour &       6,057&        6.47\\
other       &       5,746&        6.14\\
cassava     &       5,576&        5.96\\
fish        &       3,820&        4.08\\
onion       &       3,792&        4.05\\
sugar       &       2,653&        2.83\\
rape        &       2,095&        2.24\\
groundnut   &       1,928&        2.06\\
sour milk   &       1,501&        1.60\\
cowpea      &       1,316&        1.41\\
cassava leaves&       1,073&        1.15\\
sweet potato&       1,019&        1.09\\
pumpkin leaves&       1,006&        1.07\\
amaranth    &         983&        1.05\\
hibiscus    &         942&        1.01\\
rice        &         864&        0.92\\
meat        &         826&        0.88\\
tea leaves  &         817&        0.87\\
pumpkin     &         752&        0.80\\
okra        &         566&        0.60\\
\midrule Total       &       93,606&      100 \\ 
\hline \hline \\[-1.8ex] \multicolumn{3}{J{\linewidth}}{\small 
\noindent \textit{Note}: The table displays the number of times 
the top 25 ingredient was recorded in the food diaries and the relative 
frequency of that ingredient in the entire data set. In total 
111 different ingredients were recorded, excluding water.}  \end{tabular}

%% file: tables/fg_tab.tex
\begin{tabular}{l*{2}{c}} \\ [-1.8ex]\hline \hline \\[-1.8ex] 
            &   Frequency&     Percent\\
\midrule
Cereals     &      26,164&       27.95\\
Vegetables  &      21,001&       22.44\\
Spices, Condiments, \& Beverages&      18,156&       19.40\\
Oils \& Fats &      10,068&       10.76\\
Tubers      &       6,656&        7.11\\
Legumes, Nuts, \& Seeds&       4,153&        4.44\\
Fish        &       3,820&        4.08\\
Milk        &       1,800&        1.92\\
Meat        &       1,123&        1.20\\
Eggs        &         329&        0.35\\
Sweets      &         190&        0.20\\
Fruits      &         146&        0.16\\
\midrule Total       &       93,606&      100 \\ 
\hline \hline \\[-1.8ex] \multicolumn{3}{J{\linewidth}}{\small 
\noindent \textit{Note}: The table displays the number of times 
a food group is represented in the food diaries. There are 12 
total food groups.}  \end{tabular}

%% file: tables/sci_tab.tex
\begin{tabular}{l*{2}{c}} \\ [-1.8ex]\hline \hline \\[-1.8ex] 
            &   Frequency&     Percent\\
\midrule
zea\_mays    &      24,923&       37.61\\
lycopersicon\_esculentum&       7,977&       12.04\\
manihot\_esculenta&       6,661&       10.05\\
fish\_spp    &       3,820&        5.76\\
allium\_cepa\_l&       3,792&        5.72\\
bos\_taurus\_linnaeus\_1758&       2,662&        4.02\\
cucurbuta\_pepo&       2,192&        3.31\\
brassica\_rapa&       2,095&        3.16\\
arachis hypogaea&       1,928&        2.91\\
other       &       1,336&        2.02\\
vigna\_unguiculata\_l\_walp&       1,320&        1.99\\
ipomoea\_batatas&       1,318&        1.99\\
amaranthus\_spp&         983&        1.48\\
hibiscus\_acetosella&         942&        1.42\\
oryza\_sativa\_l&         864&        1.30\\
aromatic\_spp&         817&        1.23\\
gallus\_gallus\_linnaeus\_1758&         570&        0.86\\
abelmoschus\_esculentus&         568&        0.86\\
lactarius\_spp&         374&        0.56\\
phaseolus\_vulgaris\_l&         291&        0.44\\
cenchrus\_americanus&         268&        0.40\\
schinziophyton\_rautanenii&         202&        0.30\\
brassica\_oleracea\_capitata&         182&        0.27\\
cucurbita\_moschata&         177&        0.27\\
\midrule Total       &       81,330&      100 \\ 
\hline \hline \\[-1.8ex] \multicolumn{3}{J{\linewidth}}{\small 
\noindent \textit{Note}: The table displays the number of times 
a species is represented in the food diaries. There are 63 
unique species total.}  \end{tabular}

%% file: tables/out_tab.tex
\begin{tabular}{l*{5}{c}} \\ [-1.8ex]\hline \hline \\[-1.8ex] 
                    &        Dish         &        Meal         &         Day         &        Week         &       Total         \\
\midrule
Solar Stove Used    &       0.180         &       0.175         &       0.176         &       0.178         &       0.176         \\
                    &     (0.384)         &     (0.356)         &     (0.297)         &     (0.272)         &     (0.252)         \\
HDDS                &       2.317         &       3.883         &       5.693         &       8.201         &       9.833         \\
                    &     (0.999)         &     (1.539)         &     (1.477)         &     (1.504)         &     (1.212)         \\
Species Richness    &       2.321         &       3.992         &       6.676         &       13.41         &       20.15         \\
                    &     (1.020)         &     (1.754)         &     (2.294)         &     (3.806)         &     (4.734)         \\
Legumes Cooked      &       0.134         &       0.256         &       0.676         &       4.455         &       26.04         \\
                    &     (0.341)         &     (0.460)         &     (0.859)         &     (3.641)         &     (17.77)         \\
\midrule \multicolumn{1}{l}{Total} &  30,314 & 15,896 & 
 6,013 & 912 & 156\\ 
\hline \hline \\[-1.8ex] 
\multicolumn{6}{J{\linewidth}}{\small 
\noindent \textit{Note}: The table displays means and standard deviations, 
in parentheses, treatment assignment, use of solar stoves, 
and continuous control variables.}  \end{tabular}

%% file: tables/cook_tab.tex
\begin{tabular}{l*{3}{c}} \\ [-1.8ex]\hline \hline \\[-1.8ex] 
                    &         Day         &        Week         &       Total         \\
\midrule
Dishes per Meal     &       5.041         &       33.24         &       194.3         \\
                    &     (1.586)         &     (10.38)         &     (58.68)         \\
Meals Skipped       &       0.357         &       3.573         &       24.12         \\
                    &     (0.553)         &     (3.914)         &     (22.85)         \\
\midrule \multicolumn{1}{l}{Total} &  6,013 & 912 & 
 156 \\ 
\hline \hline \\[-1.8ex] 
\multicolumn{4}{J{\linewidth}}{\small 
\noindent \textit{Note}: The table displays means and standard deviations, 
in parentheses, for dishes per meal and meals skipped.}  \end{tabular}

%% file: tables/fuel_tab.tex
\begin{tabular}{l*{2}{c}} \\ [-1.8ex]\hline \hline \\[-1.8ex] 
                    &        Week         &       Total         \\
\midrule
Firewood Time (min) &       115.1         &       129.2         \\
                    &     (194.6)         &     (213.2)         \\
Firewood Quantity (kg)&       69.25         &       74.86         \\
                    &     (76.37)         &     (92.11)         \\
Charcoal Quantity (kg)&       6.285         &       5.389         \\
                    &     (19.37)         &     (15.66)         \\
Fuel Value (USD)    &       9.757         &       10.38         \\
                    &     (9.507)         &     (11.59)         \\
\midrule \multicolumn{1}{l}{Total} &  870 & 157 \\ 
\hline \hline \\[-1.8ex] 
\multicolumn{3}{J{\linewidth}}{\small 
\noindent \textit{Note}: The table displays means and standard deviations, 
in parentheses, for collected and purchased fuel.}  \end{tabular}

%% file: tables/convars_tab.tex
\begin{tabular}{l*{4}{c}} \\ [-1.8ex]\hline \hline \\[-1.8ex] 
                    &        Mean&    St. Dev.&         Min&         Max\\
\midrule
Age of Head of Household&       47.84&       12.74&          16&          80\\
Household Size      &       6.928&       2.573&           1&          15\\
Tropical Livestock Index&       5.105&       12.40&           0&       87\\
Asset Index         &       9.846&       11.46&           0&          62\\
Cloud Cover         &       99.36&       8.006&           0&         100\\
\midrule \multicolumn{4}{l}{Total Households} &      156 \\ 
\hline \hline \\[-1.8ex] \multicolumn{5}{J{\linewidth}}{\small 
\noindent \textit{Note}: The table displays summar statistics for 
treatment assignment, use of solar stoves, and continuous control variables.}  \end{tabular}

%% file: tables/catvars_tab.tex
\begin{tabular}{l*{2}{c}} \\ [-1.8ex]\hline \hline \\[-1.8ex] 
            &   Frequency&     Percent\\
\cline{2-3} \\[-1.8ex] \multicolumn{3}{l}{\emph{Treatment Assignment}} \\ 
Control     &          95&       60.90\\
Treatment   &          61&       39.10\\
\multicolumn{3}{l}{\emph{AAS Group}} \\ 
Nutrition Club&          56&       35.90\\
Both        &          54&       34.62\\
None        &          37&       23.72\\
Learning Plot&           9&        5.77\\
\multicolumn{3}{l}{\emph{Gender of Head of Household}} \\ 
Women       &         109&       69.87\\
Men         &          47&       30.13\\
\multicolumn{3}{l}{\emph{Education of Head of Household}} \\ 
Secondary   &          75&       48.08\\
Primary     &          69&       44.23\\
None        &           6&        3.85\\
Higher      &           6&        3.85\\
\multicolumn{3}{l}{\emph{Village}} \\ 
Village N   &          63&       40.38\\
Village L   &          57&       36.54\\
Village M   &          36&       23.08\\
\midrule \multicolumn{2}{l}{Total Households} &      156 \\ 
\hline \hline \\[-1.8ex] \multicolumn{3}{J{\linewidth}}{\small 
\noindent \textit{Note}: The table displays the number and 
frequency of each categorical control variable.}  \end{tabular}

%% file: tables/late_out.tex
\begin{tabular}{l*{10}{c}} \\[-1.8ex]\hline \hline \\[-1.8ex] 
& \multicolumn{2}{c}{Dish} & \multicolumn{2}{c}{Meal} & \multicolumn{2}{c}{Day} 
& \multicolumn{2}{c}{Week} & \multicolumn{2}{c}{Overall} \\ \cline{2-3} 
\cline{4-5} \cline{6-7} \cline{8-9} \cline{10-11} \\[-1.8ex] 
& \multicolumn{1}{c}{(1)} & \multicolumn{1}{c}{(2)} & \multicolumn{1}{c}{(3)} 
& \multicolumn{1}{c}{(4)} &\multicolumn{1}{c}{(5)} & \multicolumn{1}{c}{(6)} 
& \multicolumn{1}{c}{(7)} & \multicolumn{1}{c}{(8)} & \multicolumn{1}{c}{(9)} 
& \multicolumn{1}{c}{(10)} \\ \midrule 
\multicolumn{11}{l}{\emph{Panel A: Household Dietary Diversity Score}} \\ 
\midrule
Solar Stove Use     &      -0.148         &      -0.122         &      -0.279         &      -0.233         &      -0.189         &      -0.254         &      -0.221         &      -0.289         &      -0.144         &      -0.246         \\
                    &     (0.145)         &     (0.138)         &     (0.263)         &     (0.261)         &     (0.339)         &     (0.340)         &     (0.393)         &     (0.402)         &     (0.416)         &     (0.425)         \\
\midrule
Mean in Control     &       2.332         &       2.332         &       3.908         &       3.908         &       5.668         &       5.668         &       8.168         &       8.168         &       9.768         &       9.768         \\
Observations        &      30,314         &      29,707         &      15,896         &      15,578         &       6,013         &       5,891         &         912         &         894         &         156         &         153         \\
Covariates          &          No         &         Yes         &          No         &         Yes         &          No         &         Yes         &          No         &         Yes         &          No         &         Yes         \\
Adjusted R$^2$      &       0.018         &       0.027         &       0.033         &       0.042         &       0.124         &       0.128         &       0.138         &       0.140         &       0.156         &       0.139         \\
\midrule \multicolumn{11}{l}{\emph{Panel B: Dietary Species Richness}} \\ 
\midrule
Solar Stove Use     &      -0.154         &      -0.166         &      -0.188         &      -0.198         &       0.067         &      -0.071         &      -0.365         &      -0.592         &      -0.733         &      -1.007         \\
                    &     (0.163)         &     (0.160)         &     (0.326)         &     (0.322)         &     (0.567)         &     (0.558)         &     (1.166)         &     (1.177)         &     (1.813)         &     (1.874)         \\
\midrule
Mean in Control     &       2.329         &       2.329         &       3.984         &       3.984         &       6.552         &       6.552         &      13.28         &      13.28         &      20.14         &      20.14         \\
Observations        &      30,314         &      29,707         &      15,896         &      15,578         &       6,013         &       5,891         &         912         &         894         &         156         &         153         \\
Covariates          &          No         &         Yes         &          No         &         Yes         &          No         &         Yes         &          No         &         Yes         &          No         &         Yes         \\
Adjusted R$^2$      &       0.021         &       0.023         &       0.049         &       0.054         &       0.148         &       0.155         &       0.132         &       0.143         &       0.075         &       0.077         \\
\midrule \multicolumn{11}{l}{\emph{Panel C: Count of Legume Consumption}} \\ 
\midrule
Solar Stove Use     &       0.024         &       0.016         &       0.049         &       0.037         &       0.154         &       0.112         &       0.994         &       0.700         &       5.760         &       3.919         \\
                    &     (0.030)         &     (0.029)         &     (0.061)         &     (0.059)         &     (0.165)         &     (0.157)         &     (1.112)         &     (1.047)         &     (6.692)         &     (6.328)         \\
\midrule
Mean in Control     &       0.126         &       0.126         &       0.238         &       0.238         &       0.620         &       0.620         &       4.068         &       4.068         &      23.77         &      23.77         \\
Observations        &      30,314         &      29,707         &      15,896         &      15,578         &       6,013         &       5,891         &         912         &         894         &         156         &         153         \\
Covariates          &          No         &         Yes         &          No         &         Yes         &          No         &         Yes         &          No         &         Yes         &          No         &         Yes         \\
Adjusted R$^2$      &       0.006         &       0.007         &       0.018         &       0.020         &       0.046         &       0.052         &       0.107         &       0.119         &       0.133         &       0.126         \\
\hline \hline \\[-1.8ex] \multicolumn{11}{J{\linewidth}}{\small 
\noindent \textit{Note}: Dependent variables are different measure of 
 household dietary composition. In Panel A, we use dietary diversity 
score. In Panel B, we use species richness. In Panel C, we calculate 
the number of times legumes are eaten. All regressions include two levels of strata 
fixed effects: village and Agricultural and Aquatic Systems (AAS) group. 
For regressions with more than one observation per houhold (columns 1-8), 
we calculate Liang-Zeger cluster-robust standard errors since the unit 
of randomization is the household. For regressions with only one 
observation per household (columns 9-10), we calculate Eicker-Huber-White 
(EHW) robust standard errors. Standard errors are presented in 
parentheses (*** p$<$0.001, ** p$<$0.01, * p$<$0.05).}  \end{tabular}

%% file: tables/late_numdish.tex
\begin{tabular}{l*{6}{c}} \\[-1.8ex]\hline \hline \\[-1.8ex] 
& \multicolumn{2}{c}{Day} & \multicolumn{2}{c}{Week} & 
\multicolumn{2}{c}{Overall} \\ \cline{2-3} 
\cline{4-5} \cline{6-7} \\[-1.8ex] 
& \multicolumn{1}{c}{(1)} & \multicolumn{1}{c}{(2)} & \multicolumn{1}{c}{(3)} 
& \multicolumn{1}{c}{(4)} &\multicolumn{1}{c}{(5)} & \multicolumn{1}{c}{(6)} 
\\ \midrule 
\multicolumn{7}{l}{\emph{Panel A: Number of Dishes Prepared}} \\ 
\midrule
Solar Stove Use     &       0.380         &       0.300         &       2.748         &       1.944         &      16.40         &       9.287         \\
                    &     (0.347)         &     (0.342)         &     (3.144)         &     (3.139)         &    (21.38)         &    (21.21)         \\
\midrule
Mean in Control     &       4.920         &       4.920         &      32.27         &      32.27         &     188.5         &     188.5         \\
Observations        &       6,013         &       5,891         &         912         &         894         &         156         &         153         \\
Covariates          &          No         &         Yes         &          No         &         Yes         &          No         &         Yes         \\
Adjusted R$^2$      &       0.129         &       0.137         &       0.134         &       0.145         &       0.110         &       0.127         \\
\midrule \multicolumn{7}{l}{\emph{Panel B: Number of Meals Skipped}} \\ 
\midrule
Solar Stove Use     &      -0.121         &      -0.070         &      -0.950         &      -0.496         &      -5.923         &      -1.955         \\
                    &     (0.102)         &     (0.101)         &     (1.201)         &     (1.208)         &     (8.769)         &     (8.644)         \\
\midrule
Mean in Control     &       0.396         &       0.396         &       3.924         &       3.924         &      26.24         &      26.24         \\
Observations        &       6,013         &       5,891         &         912         &         894         &         156         &         153         \\
Covariates          &          No         &         Yes         &          No         &         Yes         &          No         &         Yes         \\
Adjusted R$^2$      &       0.043         &       0.065         &       0.044         &       0.072         &       0.012         &       0.102         \\
\hline \hline \\[-1.8ex] \multicolumn{7}{J{\linewidth}}{\small 
\noindent \textit{Note}: Dependent variables are different measure of 
 frequency of cooking. In Panel A, we use the number of dishes cooked 
in a meal. In Panel B, we use the number of meals skipped. 
All regressions include two levels of strata 
fixed effects: village and Agricultural and Aquatic Systems (AAS) group. 
Eicker-Huber-White (EHW) robust standard errors. Standard errors are presented in 
parentheses (*** p$<$0.001, ** p$<$0.01, * p$<$0.05).}  \end{tabular}